\documentstyle[ApJ_cite]{mn}

\begin{document}

%BEGIN USER-DEFINED SHORTFORMS
\def\spose#1{\hbox to 0pt{#1\hss}}
\def\simlt{\mathrel{\spose{\lower 3pt\hbox{$\mathchar"218$}}
     \raise 2.0pt\hbox{$\mathchar"13C$}}}
\def\simgt{\mathrel{\spose{\lower 3pt\hbox{$\mathchar"218$}}
     \raise 2.0pt\hbox{$\mathchar"13E$}}}
\def\eg{{\rm e.g. }}
\def\ie{{\rm i.e. }}
\def\etal{{\rm et~al. }}

\def\aj{{AJ}}			
\def\araa{{ARA\&A}}		
\def\apj{{ApJ}}			
\def\apjs{{ApJS}}		
\def\apss{{Ap\&SS}}		
\def\aap{{A\&A}}		
\def\aapr{{A\&A~Rev.}}		
\def\aaps{{A\&AS}}		
\def\azh{{AZh}}			
\def\jrasc{{JRASC}}		
\def\mnras{{MNRAS}}		
\def\pasa{{PASA}}		
\def\pasp{{PASP}}		
\def\pasj{{PASJ}}		
\def\sovast{{Soviet~Ast.}}	
\def\ssr{{Space~Sci.~Rev.}}	
\def\zap{{ZAp}}			
\def\nat{{Nature}}		
\def\aplett{{Astrophys.~Lett.}}
\def\fcp{{Fund.~Cosmic~Phys.}}
\def\memsai{{Mem.~Soc.~Astron.~Italiana}}
\def\nphysa{{Nucl.~Phys.~A}}
\def\physrep{{Phys.~Rep.}}
%END USER-DEFINED SHORTFORMS

\title[Supernovae-driven wind models of elliptical galaxies]{Galactic winds 
and the photo-chemical evolution of elliptical galaxies:  the classic model 
revisited}
\author[B.K. Gibson]{B.K. Gibson$^{1,2,3}$ \\
$^1$ Mount Stromlo \& Siding Spring Observatories, Australian National 
University, Weston Creek P.O., Weston, ACT, Australia  2611 \\
$^2$ Department of Astrophysics, University of Oxford, Keble Road, Oxford, UK  
OX1 3RH \\
$^3$ Department of Geophysics \& Astronomy, University of British Columbia,
Vancouver, BC, Canada  V6T 1Z4}
\maketitle
\begin{abstract}
We consider the simultaneous chemical, photometric, and gaseous thermal energy
evolution of elliptical galaxies.  The evolution of chemical abundances
in the intracluster medium (ICM)
is set by the differing timescales for gas ejection,
via supernovae (SNe)-driven winds, 
from dwarf, normal, and giant ellipticals, and is
monitored concurrently.  Emphasis is placed upon the influence of, and
sensitivity to, the underlying stellar initial mass function (IMF), 
star formation
efficiency, supernovae Type Ia rates, supernovae remnant (SNR) dynamics,
and the most recent advances in stellar nucleosynthesis.
Unlike many previous studies, we adhere to a wide range of optical (\eg
colour-metallicity-luminosity relationship) and x-ray (\eg recent ASCA ICM
abundance measurements) observational constraints.  
IMFs biased toward high mass stars, at least during the early phases of
star formation, are implicated in order to satisfy all the observational
constraints.  
\end{abstract}


%%%%%%%%%%%%%%%%%%%%%%%%%%%%%%%%%%%%%%%%%%%%%%%%%%%%%%%%%%%%%%%%%%%%%%%%%%%%%%%

\section{Introduction}
\label{introduction}

A rich history exists in the field of elliptical galaxy
spectrochemical evolution.  It was recognised
early on (\eg \cite{L74a}) that if star formation was to proceed to completion 
in all ellipticals, then the observed trend of average metallicity with the 
depth of the galactic potential well (\ie mass-metallicity relationship) 
(\cite{B59}) would be difficult to establish.
The key to understanding this observed correlation was provided by Mathews \&
Baker (1971) but not fully appreciated until Larson (1974b).

Mathews \& Baker (1971)
postulated that much of the gas in the interstellar medium (ISM)
of ellipticals had been strongly heated by supernovae (SNe) and driven out by a
hot galactic wind once the gas thermal energy exceeded that of its gravitational
binding energy (at some time $t_{\rm GW}$), 
thereby bringing to a halt the bulk of active star formation.  Larson (1974b)
noted that the binding energy per unit mass of gas is higher in the more
massive galaxies, and thus these systems would retain their gas for a longer
period of time before reaching $t_{\rm GW}$, and thus attain a higher
metallicity, consistent with the observed mass-metallicity relationship.
The subsequent evolution would then be regulated only by gas lost from dying
stars.  An excellent recent review of galactic winds, from an observational
slant, is given by Bland-Hawthorn (1995).

What began with Larson (1974b) has led to a profusion of follow-up SNe-driven
galactic wind models (\eg \cite{I77}; \cite{S79b}; \cite{DS86};
\cite{AY87} (AY87); \cite{MT87} (MT87); \cite{AG90} (AG90); \cite{CDPR91};
\cite{DFJ91}; \cite{BR92};
\cite{ARBVV92}; \cite{FP93}; \cite{RCDP93}; \cite{OCKF93}; \cite{BCF94};
\cite{MT94}; \cite{EAV95}; \cite{NC95}; \cite{GM96}).
Modern advances in stellar nucleosynthesis, supernova remnant (SNR)
shock dynamics and thermal evolution, 
the role of dark matter, and even basic observational constraints 
(both in the stellar
component of ellipticals and the intracluster medium (ICM) of galaxy clusters) 
makes a re-examination of the classic wind model a timely one.
We have developed a software 
package, entitled {\bf MEGaW}$\equiv${\bf M}etallicity {\bf
E}volution with {\bf Ga}lactic {\bf W}inds, to enable us to study the
chemical evolution of ellipticals within the framework of Larson (1974b)
classic galactic wind model. 

Concurrent to the development of sophisticated wind models was the flourishing
field of spectral and photometric evolution of galaxies, the early history
being traced in Tinsley's (1980)
seminal paper.  One need only look to the recent
models of Bruzual \& Charlot (1993), Bressan \etal (1994), Worthey (1994) and
Einsel \etal (1995), for a few state of the art examples.
The consideration of chemical evolution together with
photometric evolution is imperative as stellar
evolutionary tracks and spectrophotometric calibrations
are sensitive to the chemical composition, and elliptical
galaxies are complex systems with a distribution of stellar populations.  
Models which utilise only solar abundance tracks and solar
abundance spectra/colours (\eg \cite{GRV87}; \cite{BC93}) 
fail to
explain the observed correlation between the integrated colours and absolute
magnitude of ellipticals (\eg \cite{F77}; \cite{BCF94}) (as well as,
obviously, the metallicity-luminosity relationship).

Before embarking upon the construction of a photo-chemical evolution code
suitable for both the elliptical galaxies in question and ICM
abundances, it is important to be aware of some of
the primary observational constraints\footnote{$H_0\equiv 50$ km/s/Mpc assumed
throughout.}:

\begin{itemize}
\item{\it Elliptical CML Relations: }
First and foremost, for the underlying ellipticals, the models
must honour the observed, present-day
colour-metallicity-luminosity (CML)
relationships (\eg \cite{F77}). \ie ellipticals,
in general, show increasing metallicity and redder colours, with increasing
luminosity.
\item{\it Magnesium Overabundance in Stellar Populations: }
[Mg/Fe] in giant ellipticals is $\sim +0.2\rightarrow
+0.3$, with a slight trend toward increasing values with increasing luminosity,
albeit with a fairly large degree of scatter (\cite{WFG92}).
\item{\it Oxygen Overabundance of the ICM: }
X-ray observations of the hot ICM of four clusters of galaxies shows an oxygen
overabundance relative to iron of [O/Fe]=$+0.4\pm 0.3$
(\cite{Mush94}). 
\item{\it ICM Iron-Luminosity Relation: }
The ICM iron mass (in $M_\odot$) increases as a function of a
cluster's optical luminosity tied up in E$+$S0s (in L$_{\rm V_\odot}$), 
such that
$M_{\rm Fe}^{\rm ICM}\approx 0.02 L_{\rm V}^{\rm E+S0}$ (\cite{ARBVV92}).
\item{\it Type Ia Supernovae Rates in Ellipticals: }
The present-day Type Ia SN rate
in giant ellipticals is $R_{\rm Ia}\approx 0.03\rightarrow 0.08$ 
SNu\footnote{1 SNu$\equiv$ 1 SN/century/$10^{10}$ L$_{{\rm B}_\odot}$.}
(\cite{TCB94}).
\end{itemize}

The work described herein is the only study to date to draw upon \it all \rm
of the above;
the previously mentioned models neglect, or fail to satisfy, one or more
of these constraints.  Photometric self-consistency is the usual constraint
which suffers, primarily due to a past dearth of non-solar
calibrations/tracks (although, see \cite{AY87} and \cite{BCF94}).  
Overcoming this major hurdle is only now becoming feasible; specifically,
the past two years has seen the release of the first wave of
fully self-consistent isochrones, ranging from ultra-low to super metal-rich,
isochrones.  These isochrone compilations (\cite{W94} and \cite{BBCFN94})
draw heavily upon Kurucz's (1993) extensive grid of model stellar atmospheres.  

Section \ref{model} introduces the galactic wind framework in which {\bf MEGaW}
operates.  This includes an outline of the relevant chemical evolution
equations, as well as a description of the different parametrisations
adopted for each of the input ingredients.  In Section \ref{template} we
present a template model against which future comparisons will be made.
Sections \ref{analtau} to \ref{analsfr} then systematically explore
the sensitivity of the model predictions to the different input ingredients.
Such an analysis is long overdue, as previous studies tended to 
simply select one preferred combination
of parameters without illustrating the ramifications of said selection.  
A summary of the primary conclusions can be found in Section \ref{summary}.

%%%%%%%%%%%%%%%%%%%%%%%%%%%%%%%%%%%%%%%%%%%%%%%%%%%%%%%%%%%%%%%%%%%%%%%%%%%%%%%
\section{The classic galactic wind model}
\label{model}
\subsection{Supernova progenitors and rates}
\label{progenitors}
We adopt the ``single degenerate'' model of \cite{WI73}
for Type Ia SNe in which the
progenitors are C/O-white dwarfs in close
binary systems accreting material from a secondary companion.  
Type II SNe are presumed to originate via the core bounce-induced explosion
of single massive star (\ie $m\simgt 8$ M$_\odot$) progenitors.  Arnett (1996)
provides an excellent review of the relevant SN physics.

As in Matteucci \& Tornamb\`e (1987), we have chosen to use the SN rate
formalisms proposed by Greggio \& Renzini (1983). 
Calling $m_1$ and $m_2$ the mass of the primary and
secondary, respectively
(\ie $m_{\rm B}=m_1+m_2$), and denoting the secondary mass fraction $\mu\equiv
m_2/m_{\rm B}$, we can write the Type Ia SN rate 
$R_{\rm Ia}(t)$ as: 
\begin{equation}
R_{\rm Ia}(t) = 
A\int_{m^1}^{m_{\rm BM}}
{{\phi(m)}\over{m}}
\biggl\{\int_{\mu_{\rm m}}^{0.5}f(\mu)\psi(t-\tau_{m_2}){\rm d}\mu\biggr\}
{\rm d}m
\label{eq:Ia_rate}
\end{equation}
\noindent
where $m^1={\rm max}[2m_t,3.0]$.  For a given $t$,
$m_t$ represents the mass of stars currently
leaving the main sequence.
$\phi(m)$ is the IMF, by mass.
The distribution function for the mass fraction
of secondaries is taken to be proportional to $\mu^2$, and
the minimum mass fraction $\mu_{\rm m}$
contributing to the SN Type Ia rate at time $t$ is $\mu_{\rm m}\equiv
{\rm max}\big[m_2(t)/m_{\rm B},1-0.5m_{\rm BM}/m_{\rm B}\big]$ 
(\cite{GR83}).  $A$ represents the 
mass fraction of the IMF which is tied up in
binary systems with total masses in the range 
$m_{\rm Bm}$(3.0 ${\rm M}_\odot)\rightarrow m_{\rm BM}$(16.0 ${\rm M}_\odot)$.
$A$ is fixed \it a posteriori \rm by ensuring
reproduction of the current rate of Type Ia events in ellipticals.

The Type II SN rate $R_{\rm II}(t)$ 
is composed of two terms -- one for all stars with initial
masses greater than 16.0 M$_\odot$, and one for the fraction (\ie $1-A$) 
of stars in the mass range $8.0\rightarrow 16.0$ M$_\odot$ which are not part
of binary systems:
\begin{eqnarray}
R_{\rm II}(t) & = & (1-A)\int_{{\rm max}[m_t,8.0]}^{m_{\rm BM}}{{\phi(m)}
\over{m}}\psi(t-\tau_m){\rm d}m\,\, + \nonumber \\
& & \int_{m_{\rm BM}}^{m_{\rm U}}
{{\phi(m)}\over{m}}\psi(t-\tau_m){\rm d}m. \label{eq:II_rate}
\end{eqnarray}

Single low and intermediate mass stars ($m\simlt 8.0$ M$_\odot$) are presumed to
end their lives as white dwarfs, after passing through a thermally
pulsing-asymptotic giant branch (TP-AGB) and planetary nebulae (PNe) ejection
phase (\cite{RV81}).

\subsection{Stellar lifetimes}
\label{ingredtau}
Stars return gas and metals to the ISM via stellar winds, SNe events and
envelope ejection, depending upon the initial mass and metallicity.
The bulk of the ejection occurs near the end of the star's lifetime
$\tau(m,{\rm Z})$, with stellar evolutionary theory providing the functional
form of $\tau$.  

Figure \ref{fig:tau} illustrates three popular compilations of
$\tau$ versus inital mass $m$:  AY87 and AG90 adopted the singular power-law
form from Talbot \& Arnett (1971), 
which predicts excessively long lifetimes for stars in the
$\sim 1\rightarrow \sim 50$ M$_\odot$ range; 
MT87 used G\"usten \& Mezger's (1982) parametrization of the Alcock \&
Paczynski (1978) and Becker (1981)
stellar models\footnote{Because Alcock \& Paczynski (1978) and
Becker (1981) did not extend their stellar evolution studies below $m=2$
M$_\odot$, nor consider metallicities besides Z=0.03, G\"usten \& Mezger (1982)
were forced
to extrapolate into regimes the models were not designed to replicate.  This
results in considerably shorter lifetimes for solar and sub-solar mass stars,
in comparison with present-day models.}.  The best approach available today is
to adopt the metallicity-dependent lifetimes provided
by newer, more extensive, 
grids of stellar evolution tracks (\eg the Padova Group -- \cite{FBBC94}
or the Geneva Group -- \cite{SSMM92}).  Unless otherwise noted,
we use the lifetimes from Schaller \etal's (1992)
Z=0.02 grid, although for
lower metallicity systems, their Z=0.001 predictions are adopted.

Varying the selection of stellar lifetime formalism is considered in Section
\ref{analtau}.

\begin{figure}[ht]
\centering
\vspace{8.5cm}
\includegraphics{`cat}
\caption[]{Stellar lifetimes $\tau$ as a function of
initial mass $m$.  
TA71=Talbot \& Arnett (1971); GM82=G\"usten \& Mezger (1992); 
SSMM92=Schaller \etal (1992).
\label{fig:tau}}
\end{figure}

\subsection{Chemical evolution}
\label{ingredeqns}
Models for chemical evolution follow abundance changes in the ISM
of a region, and the resulting abundance distributions in stars.  A
detailed derivation of the fundamental equations can be found in Talbot \&
Arnett (1971) and Tinsley (1980).
We have chosen to couple
these equations (appropriate for the one-zone model
with instantaneous mixing of gas and relaxation of the instantaneous recycling
approximation) with the classic galactic wind model 
formalism of Matteucci \& Tornamb\`e (1987).

The mass of the ISM (\ie gas mass $M_{\rm g}(t)$) changes through star formation
and ejection:
\begin{equation}
{{{\rm d}M_{\rm g}(t)}\over{{\rm d}t}} = -\psi(t) + E(t), \label{eq:gas}
\end{equation}
\noindent
where $\psi(t)$ is the mass of gas being converted into stars per unit time:
\begin{equation}
\psi(t) = \nu M_{\rm g}(t),     \label{eq:sfr}
\end{equation}
\noindent
and $E(t)$ is the total 
ejection rate of gas from all stars:
\begin{eqnarray}
\lefteqn{E(t) = \int_{m^1}^{m_{\rm Bm}}\phi(m)
\psi(t-\tau_m)R(m){\rm d}m+ \nonumber } \\
 & & A\int_{m^2}^{m_{\rm BM}}\phi(m)
\Biggl\{\int_{\mu_m}^{0.5}f(\mu)\psi(t-\tau_{m_2})
R(m_2){\rm d}\mu\Biggr\}{\rm d}m+ \nonumber \\
 & & (1-A)\int_{m^2}^{m_{\rm BM}}\phi(m)
\psi(t-\tau_m)R(m){\rm d}m+ \nonumber \\
 & & \int_{m^3}^{m_{\rm U}}\phi(m)
\psi(t-\tau_m)R(m){\rm d}m.   \label{eq:gasejectrate}
\end{eqnarray}
\noindent
The star formation rate denoted by equation \ref{eq:sfr} is equivalent to
assuming a Schmidt (1959) Law with exponent one.
The respective integration lower limits to equation \ref{eq:gasejectrate} are
$m^1 = {\rm max}[m_t,m_{\rm L}]$, 
$m^2 = {\rm max}[m_t,m_{\rm Bm}]$, and
$m^3 = {\rm max}[m_t,m_{\rm BM}]$.
$R(m)=[1-w_m/m]$ represents the fractional mass of a star of 
initial mass $m$ and remnant mass $w_m$,
ejected back into the ISM after its stellar lifetime $\tau$.
The lifetime $\tau$ and remnant mass $w_m$ both depend upon a
star's metallicity Z.  Unless otherwise noted, we use the remnant mass
formalism of Prantzos \etal (1993).
Following the majority of the earlier studies, we take the star formation rate
$\psi$ during the post-galactic wind phase to be zero\footnote{For the 
broad-band colours to which the current code is aimed at reproducing
(\eg B-V, V-K, etc), relaxing such a restriction does not dramatically alter the
model results;
Arimoto (1989), Ferrini \& Poggianti (1993), and Elbaz \etal (1995), each
provide evidence for the limited importance of post-$t_{\rm GW}$ star
formation.  On the other hand, such an assumption is most likely \it not \rm
an ideal one when one is more 
concerned with replicating specific line indices (\eg \cite{WTF95}); because we
are still in the process of extending \bf MEGaW\rm's functionality to 
include full spectral synthesis, we have chosen the more conservative
$\psi(t>t_{\rm GW})\equiv 0$ route, for the time being.}.

The first integral on the right-hand side of equation \ref{eq:gasejectrate}
represents the rate at which gas is being returned to the ISM at time $t$ from
single low mass ($m\simlt 8$ M$_\odot$) stars ending their lives as white
dwarfs.  Integral two is the gas mass return rate from stars in binary systems
which end their lives as Type Ia SNe.  
Integral three is the rate of gas mass ejection at time $t$ from single stars in
the mass range $3.0\rightarrow 16.0$ M$_\odot$.  For $m<8.0$ M$_\odot$, these
single stars end as white dwarfs, whereas for $8.0<m<16.0$ M$_\odot$ they end
as Type II SNe. 
The final integral is the gas ejection 
rate from single massive stars (\ie 16.0 M$_\odot$
$\rightarrow m_{\rm U}$) which end their lives as Type II SNe.

The equation governing the evolution of the mass of metals 
$M_{\rm Z}(t)\equiv M_{\rm g}(t){\rm Z}(t)$ in the ISM gas is
\begin{equation}
{{{\rm d}M_{\rm Z}(t)}\over{{\rm d}t}} = -{\rm Z}(t)\psi(t) + E_{\rm Z}(t),
\label{eq:metals}
\end{equation}
\noindent
where $E_{\rm Z}(t)$ is the total ejection rate of new and old metals (\ie
processed and unprocessed, respectively).  The appropriate form for $E_{\rm
Z}(t)$ can be recovered by replacing $mR(m)$ in equation \ref{eq:gasejectrate}
with $m_{{\rm Z},m}^{\rm ej}$ (Section \ref{ingredyields}), 
the total mass of metals ejected from a star of
initial mass $m$ and initial metallicity Z($t-\tau_m$).

As a final sanity check on our numerical solutions of equations \ref{eq:gas}
and \ref{eq:metals}, we were fortunate to find several experts in the field
willing to run well-defined standard 
models with which to compare with our own code --
specifically, Francesca Matteucci (SISSA/Trieste - \cite{M92}), Frank Timmes 
(Chicago - \cite{TWW95}) and Leticia Carigi (CIDA - \cite{C94}).
Disregarding minor differences incurred by the various approaches to 
nucleosynthesis in the highly
uncertain $m\sim 8\rightarrow 11$ M$_\odot$ range, and assumptions regarding
the fate of unprocessed metals in low mass star ejecta, the results of the
intercomparison were more than satisfactory, 
and put this author's mind at ease.  The biggest single discrepancy occurred in
the magnesium
evolution, with Matteucci's code predicting lower values,
a result which was anticipated, and simply reflects 
a difference in the adopted Mg yield between the two codes, and not a
numerical problem.  We return to this difference in subsequent sections.

\subsection{Stellar nucleosynthesis yields}
\label{ingredyields}

The key ingredient in any chemical evolution code will obviously be the adopted
nucleosynthetic stellar yields (\ie the mass of element $i$ ejected by a star
of initial mass $m$ during the course of its lifetime).  These yields, as
provided by practitioners of stellar evolution and nucleosynthesis, are in
general parametrised in terms of a progenitor's initial mass $m$, 
metallicity Z, and ``environment'' (\ie single or binary system).  

Type II
SNe are responsible for producing the bulk of heavy elements;
near the end of the progenitor's (Section
\ref{progenitors}) lifetime, during the
carbon-burning phase, a He-exhausted core (primarily iron-peak nuclei)
becomes effectively isolated from the rest of the star.  The subsequent core
collapse leads to $\sim 10^{51}$ erg being deposited into
the overlying mantle (neutrino energy), the final result being
a compact remnant (neutron star or black hole) and an ejected envelope,
enriched in metals.

The last few years have seen an explosion of interest in massive star (\ie
$m\simgt 10$ M$_\odot$) evolution, the result being the newly (or
soon-to-be) available metallicity-dependent yields of Maeder (1992, hereafter
M92), Woosley \& Weaver (1995, hereafter WW95), and Langer \& Henkel (1995,
hereafter LH95).
Improved solar metallicity tabulations from Arnett (1991, hereafter A91) and
Thielemann \etal (1995, hereafter TNH95)
have also come on-line.  Each of these five
options are available in {\bf MEGaW}.

Each of the compilations start at the approximate lower mass cut-off for
Type II SNe (\ie $\sim 10\rightarrow 
12$ M$_\odot$), but have very different upper mass
limits.  M92 goes as high as 120 M$_\odot$ for both Z=0.001 and
Z=0.020; A91 goes to 85 M$_\odot$, but only for Z=0.020; LH95
have an upper limit of 50 M$_\odot$ for their models, with 
Z=0.002 and Z=0.020; WW95 have the best metallicity coverage
(Z=0.0000, 0.00002, 0.0002, 0.0020, and 0.0200), with an upper limit of 40
M$_\odot$; TNH95 only go as high as 25 M$_\odot$, and only for Z=0.02.
We mention in passing that WW95 consider three different models for 
$m\ge 30$ M$_\odot$, their so-called A, B, and C models.  These differ in the
amount of energy imparted by the piston in their models at explosion
initiation.
Following Timmes \etal (1995), we use the ``B'' models in this mass regime.

While each source provides the yields for H, He, Z, C, and O, the remaining
elements under consideration here are covered by some, but not by others.  N is
listed by all except A91, while Mg, Si, and Fe were not part of the
M92 or LH95 tables.  For the sake of self-consistency with the
stellar evolution models, we do not attempt to ``mix-and-match'' in order to
``fill-in'' those elements missing in one compilation with those found in
another, as the input physics between models is usually incompatible from one
to another.  The mix-and-match approach is a necessary one for those using the
M92 yields, as is seen in Carigi (1994) and Giovagnoli \& Tosi (1995) -- \eg
for the evolution of Fe, the former adopts a yield of 0.075 M$_\odot$
for all Type II SNe, irrespective of progenitor mass or metallicity, while the
latter adopts the Fe yields of Arnett (1978).

A detailed comparison of the five Type II SNe
yield options will be published elsewhere (Gibson 1997, in preparation). This
will include a comprehensive outline of the different assumptions regarding the
input physics.  We
refer the reader to each of the relevant references for tabulated values of
$m^{\rm ej}_{i,m}$ (equation \ref{eq:metals}), although we do draw attention
here to several interesting points which can be made from a 
cursory glance at some of the predicted yield ratios from said compilations.

\begin{itemize} 
\item{\it [O/Fe]: } Mushotzky (1994) has demonstrated
that the intracluster media (ICMs) of the
four clusters in their ASCA analysis, is overabundant in oxygen
when compared to iron, by factors of $\sim 2\rightarrow 5$, with respect to the
solar ratio.  This observation is reflected by the shaded region of Figure
\ref{fig:mf89_ofe}.  Contrast this constraint with the predicted [O/Fe] yield
ratios from the three compilations which include iron (\ie WW95,TNH95,A91) -- 
it is readily apparent
that galactic ejecta enriched in the byproducts of Type II SNe of mass
$m\simgt 20$ M$_\odot$ are the \it necessary \rm primary\footnote{Because
the other primary source of heavy
element enrichment -- \ie Type Ia SNe -- have
[O/Fe]$\approx -1.53$ (\cite{TNH93}), they can only
act to drive [O/Fe] downward from the observed overabundance.  
As Type Ia SNe
tend to dominate the enrichment history beyond $t\simgt 1$ Gyr (\cite{MG86};
Section \ref{template}), 
this ICM observation provides indirect evidence that early,
Type II SNe-driven winds are the dominant mechanism at play.}
contributor of the bulk of these heavy elements in the ICM.  The yield ratios
adopted in Matteucci's (1992) chemical evolution code are indicated by the
dotted line in Figure \ref{fig:mf89_ofe}.\footnote{Matteucci (1992) assumed
a constant oxygen yield of $\sim 0.48$ M$_\odot$ for all stars of initial mass
$9\simlt m\simlt 15$ M$_\odot$, suppressing the ejection of all other heavy
elements.  On the other hand, following Iwamoto \etal (1994), we have chosen to
suppress the ejection of \it all \rm newly synthesised metals in the
$8\rightarrow 11$ M$_\odot$ range.}
For $m\simgt 30$ M$_\odot$, 
A91's different mass cut results in more iron being ejected
for a given initial mass (Gibson 1997, in preparation).  TNH95's
significantly coarser grid makes a detailed comparison difficult, and simple
extrapolation to higher masses was assumed.
\item{\it [Mg/Fe]: } The stellar populations of elliptical galaxies posess
a $\sim 50\rightarrow 100$\% overabundance of magnesium-to-iron, in
comparison with the solar ratio (see the shaded region of Figure
\ref{fig:mf89_mgfe} -- Worthey \etal 1992).  Considering the oxygen
overabundance relative to iron seen in cluster ICMs, with the fact that the
bulk of star formation in the contributing ellipticals ceases subsequent to the
galactic wind epoch, this should not be entirely unexpected.\footnote{Type Ia
SNe contribute Mg and Fe in the ratio [Mg/Fe]$\approx
-1.63$ (\cite{TNH93}).  Similar to the ICM [O/Fe], this lends support to the
notion that a Type II SNe-driven wind, as opposed to a later Type Ia-driven
one, is dominating the evolution.}  Significant differences in the behaviour of
[Mg/Fe] appear to exist in the stellar models in question.
Factors of three to ten between 
A91, WW95, and TNH95, at a given mass, are apparent.  The WW95
curve lies 
consistently below A91 and TNH95 for $m<35$ M$_\odot$.  These obvious
differences must be borne in mind when attempting to replicate Worthey \etal's
(1992) observation of [Mg/Fe]$\approx +0.2\rightarrow +0.3$ in ellipticals.
The dotted line in Figure \ref{fig:mf89_mgfe} shows the adopted
[Mg/Fe] in Matteucci's (1992) code.\footnote{
Matteucci's [Mg/Fe] (Figure \ref{fig:mf89_mgfe}) is fairly flat,
and, more importantly, negative for all $m\simlt 35$ M$_\odot$.  This would
appear to be due to a lower adopted magnesium yield in her code, which was
based upon the state-of-the-art during the mid-1980s.
Recalling that the [Mg/Fe] of the stellar component of giant ellipticals
is typically $+0.2\rightarrow +0.3$, one can anticipate that fairly extreme
IMFs, with a substantial $m\simgt 35$ M$_\odot$ component,
would be \it necessary \rm with her code, in order to recover
such super-solar Mg overabundances.  This is indeed what Matteucci (1994)
found.}
\end{itemize} 

\begin{figure}[ht]
\centering
\vspace{8.5cm}
\includegraphics{`cat}
\caption[]{Comparison of the Type II SNe ejecta [O/Fe] from 
three of the sources
discussed in Section \ref{ingredyields}.  The shaded region represents the
observed ICM [O/Fe] (Mushotzky 1994).  Also shown is the yield ratio adopted in
Matteucci's (1992) chemical evolution code.
\label{fig:mf89_ofe}}
\end{figure}

\begin{figure}[ht]
\centering
\vspace{8.5cm}
\includegraphics{`cat}
\caption[]{Comparison of the Type II SNe ejecta [Mg/Fe] from 
three of the sources
discussed in Section \ref{ingredyields}.  The shaded region represents the
observed elliptical galaxy 
[Mg/Fe] (Worthey \etal 1992).  Also shown is the yield ratio adopted in
Matteucci's (1992) chemical evolution code.
\label{fig:mf89_mgfe}}
\end{figure}

The yields for single low and intermediate mass stars (\ie $m\le 8$ M$_\odot$)
are taken from Renzini \& Voli (1981, hereafter RV81).
Stars in this mass range are seen to be important contributors
to He, C, and N.  
These lower mass stars are not capable of igniting C in their cores
(their masses are too low), and end their lives as He or C/O white dwarfs,
after passing through several dredge-up phases, a thermally pulsing
phase on the upper AGB, and finally a PN ejection.  
We generally use RV81's preferred
models with the Reimers' (1975) mass loss parameter on the red giant branch
(RGB) $\eta=0.33$,
and the ratio of the mixing length to
the pressure scale height $\alpha=1.5$.\footnote{Specifically, 
Tables 3d and 3i (Z=0.020 and 0.004, respectively) from RV81 for 
$m\ge 4$ M$_\odot$, and Tables 3a and 3h (Z=0.020 and 0.004, respectively) for
$m<4$ M$_\odot$, were adopted.  The RV81 models are admittedly outdated, and
our code has subsequently been updated to accommodate RV81, as well as the newer
Marigo \etal (1996) and van den Hoek \& Groenewegen 
(1997) compilations.  None of
the results presented herein depend upon the selection of low mass stellar
yields, and we have simply retained RV81 for the discussion which follows.  A
companion paper (Gibson \& Mould 1997) contrasts these
compilations, as pertaining to chemical evolution of the Galactic halo.}

Stars with initial masses between $\sim 8$ and $\sim 11$ M$_\odot$ undergo
non-degenerate C-burning and develop O+Ne+Mg cores.  Whether these highly
enriched cores contribute their newly synthesised metals to the ISM via
thermonuclear explosion, or else undergo core collapse, trapping the yields in
the remnant neutron star, is still unclear (\eg \cite{M91}).  
Much depends upon the
assumed electron capture onto $^{24}$Mg and $^{20}$Ne, but the controversy does
seem to be converging toward the picture in which core collapse is favoured,
and the elemental enrichment is basically restricted to that deposited during
the pre-SN stellar wind (\cite{INH94}), and this is what has been assumed in
{\bf MEGaW}.  For comparison, Matteucci (1992) suppresses all newly synthesised
metals except oxygen,
and Timmes \etal (1995) simply interpolate
linearly between the highest mass in RV81 (\ie 8 M$_\odot$) and the
lowest mass in WW95 (\ie $\sim 11$ M$_\odot$). 

The yields for each Type Ia event are taken from the improved Z=0.02 W7 Model 
presented in Thielemann \etal (1993).
Of the $\sim 1.36$ M$_\odot$ of metals ejected per event,
iron is by far the largest contributor ($\sim$ 0.74$ M_\odot$), with lesser
amounts being ejected in the form of Si, O, C, and Mg (\ie 0.15, 0.14, 0.05,
and 0.01 M$_\odot$, respectively.
These yields are assumed to be
metallicity-independent -- a reasonable assumption based upon the similarity of
Thielemann \etal's (1993) Z=0.02 and Z=0.00 models.

Unless otherwise noted, the default yields adopted throughout this paper are
culled from WW95 ($m\simgt 10$ M$_\odot$, RV81 ($m\simlt 8$ M$_\odot$), and
Thielemann \etal (1993) (for Type Ia SNe).
How the predicted
chemical evolution is affected by a choice of yield tables different from that
of WW95 is discussed further in Section \ref{analyields}, and in particular, by
Gibson (1997, in preparation).  

\subsection{Thermal and binding energy evolution of the ISM}
\label{ingredenergy}
\subsubsection{ISM thermal energy}
For gas to be expelled from a galaxy, the thermal energy of the gas heated by 
SN explosions and stellar mass loss must
overcome the binding energy of the gas (\cite{L74b}) --
\ie the galactic wind will
start at a certain time $t_{\rm GW}$ when
\begin{eqnarray}
E_{\rm th}(t_{\rm GW}) & \equiv & E_{{\rm th}_{\rm Ia}}(t_{\rm GW}) + 
E_{{\rm th}_{\rm II}}(t_{\rm GW}) + E_{{\rm th}_{\rm W}}(t_{\rm GW}) \nonumber
\\
& = & \Omega_{\rm g}(t_{\rm GW}), \label{eq:windcondition}
\end{eqnarray}
\noindent
where $E_{\rm th}(t)$ is the total thermal energy in the gas at the time $t$,
given by the sum of the contribution from SNe Types Ia and II 
($E_{{\rm th}_{\rm Ia}}(t)$ and $E_{{\rm th}_{\rm II}}(t)$,
respectively) and thermalised kinetic energy from mass loss in high mass 
($m\simgt 12$ M$_\odot$) stars ($E_{{\rm th}_{\rm W}}(t)$, which is given by
equation 5 of Gibson 1994a).

The SNe thermal energy components of equation \ref{eq:windcondition}
can be written (\cite{S79b}):
\begin{eqnarray}
E_{{\rm th}_{\rm Ia}}(t) & = &
\int_0^t\varepsilon_{{\rm th}_{\rm SN}}(t-t^\prime)R_{\rm Ia}(t^\prime)
{\rm d}t^\prime \\ \nonumber
E_{{\rm th}_{\rm II}}(t) & = &
\int_0^t\varepsilon_{{\rm th}_{\rm SN}}(t-t^\prime)R_{\rm II}(t^\prime)
{\rm d}t^\prime, \label{eq:snismenergy}
\end{eqnarray}
\noindent
where
$t^\prime$ is the SN explosion time, $R_{\rm Ia}$ and $R_{\rm II}$ are the
SNe Ia and II rates (equations \ref{eq:Ia_rate} and \ref{eq:II_rate}, 
respectively), and $\varepsilon_{{\rm th}_{\rm SN}}$ 
is the equation governing the
evolution of the thermal energy content in the interior of a supernova remnant
(SNR) (Section \ref{ingredsnr}).  In the post-wind phase ($t>t_{\rm GW}$), the
lower limit on the time integrals in equations \ref{eq:windcondition} 
is taken to be $t_{\rm GW}$.

\subsubsection{ISM binding energy}
In order to determine the onset of the galactic wind, we need to compute the
binding energy of the gas as a function of time, $\Omega_{\rm g}(t)$.  
As discussed in Matteucci (1992), 
$\Omega_{\rm g}(t)$ is influenced by the presence
of dark matter, and its distribution relative to the luminous
(\ie gas + stars) component.  We consider two different scenarios: (i) one in
which the two components trace each other, following the prescription of
Saito (1979b), and (ii) one in which the luminous component is embedded in a
massive, diffuse halo of dark matter, adopting the two-component, 
self-consistent models of Bertin \etal (1992).

\vskip2.0truemm
\noindent
\underline{Dark matter traces luminous component}

Following Arimoto \& Yoshii (1987), Matteucci \& Tornamb\`e (1987), and
Angeletti \& Giannone (1990), we first
represent elliptical galaxies as
spheroidal, homogeneous systems, with a characteristic
gravitational radius
$R_{\rm G}$, and total mass $M_{\rm G}\equiv M_{\rm L}+M_{\rm D}$ 
(\ie the sum of the luminous and dark components).
Assuming the virial theorem
holds, the total binding energy of the system $\Omega_{\rm G}$ can be written as
\begin{equation}
\Omega_{\rm G} = {{GM_{\rm G}^2}\over{2R_{\rm G}}}. \label{eq:be_BT}
\end{equation}
\noindent

Following Saito (1979b),
the binding energy of the gaseous component $\Omega_{\rm g}(t)$ alone can be
written
\begin{equation}
\Omega_{\rm g}(t) = \Omega_{\rm G}{{M_{\rm g}(t)}\over{M_{\rm G}}}\biggl(2-
{{M_{\rm g}(t)}\over{M_{\rm G}}}\biggr), \label{eq:gasbe}
\end{equation}
\noindent
It is this binding energy which is compared with the gaseous thermal energy
calculated from equation \ref{eq:windcondition} 
in order to determine the time of
galactic wind onset $t_{\rm GW}$.

\vskip2.0truemm
\noindent
\underline{Diffuse dark halos}

While the previous section's analysis is suitable for models in which the dark
matter component is distributed similarly to the luminous component, it is not
suitable for the more generally accepted scenario in which the gas and stars
are embedded in a massive diffuse halo of dark matter (\eg \cite{BBB94}, and
references therein).  To model such distributions, we follow the prescription
laid out in Bertin \etal (1992).
In this context, the binding
energy of gas is expressed as 
\begin{equation}
\Omega_{\rm g}(t)=\Omega_{\rm L}(t)+\Omega_{\rm LD}(t),   \label{eq:be_BSS}
\end{equation}
\noindent
where the gravitational energy of the gas due to the luminous matter
$\Omega_{\rm L}$ (dropping the implicit time dependency) is
\begin{equation}
\Omega_{\rm L} = {{GM_{\rm g}M_{\rm L}}\over{2R_{\rm L}}}  \label{eq:omega_lum}
\end{equation}
\noindent
and the gravitational energy of the gas due to the dark matter $\Omega_{\rm
LD}$ is
\begin{equation}
\Omega_{\rm LD} = {{GM_{\rm g}M_{\rm D}}\over{R_{\rm L}}}\Omega^\prime_{\rm LD}.
\label{eq:omega_dark}
\end{equation}
\noindent
The interaction integral $\Omega^\prime_{\rm LD}$ is estimated to be
\begin{equation}
\Omega^\prime_{\rm LD} \approx {{1}\over{2\pi}}{{R_{\rm L}}\over{R_{\rm D}}}
\Biggl[1+1.37\biggl({{R_{\rm L}}\over{R_{\rm D}}}\biggr)\Biggr].
\label{eq:omega_prime}
\end{equation}
\noindent
$R_{\rm D}$ and $M_{\rm D}$ is the radial extent and mass of the dark matter
halo.  $M_{\rm D}\simgt (2\rightarrow 10)M_{\rm L}$ and $R_{\rm D}\simgt
5R_{\rm L}$ are favoured in these two-component models when applied to large
samples of ellipticals (\cite{SBS92}).

\vskip2.0truemm
\noindent
\underline{Mass-energy-radius relations}

One last piece of the puzzle necessary to solve for the gaseous binding energy
is obvious upon re-inspection of equations \ref{eq:be_BT}, \ref{eq:gasbe} 
and \ref{eq:be_BSS} -- a relationship between the galaxy mass, radius, and
binding energy.  To this end, we use empirical relationships derived by
Saito (1979a) from fitting Michie-King models to the surface brightness
distribution and line-of-sight velocity dispersions for a wide range of nearby
spheroids (ranging from globular clusters to giant ellipticals).  This first
gives the $\Omega_{\rm L}$--$M_{\rm L}$ relation:
\begin{equation}
\Omega_{\rm L} = 
1.64\times 10^{60}\Biggl[{{M_{\rm L}}\over{10^{12}}}\Biggr]^{1.45}, 
\quad[{\rm erg}]
\label{eq:be_s79}
\end{equation}
\noindent
where $M_{\rm L}$ is in solar masses.  Combining equations
\ref{eq:be_BT} and \ref{eq:be_s79}, we can write the virial
radius of our spheroidal
galaxies in terms of the mass as
\begin{equation}
R_{\rm L} = 26.0\Biggl[{{M_{\rm L}}\over{10^{12}}}\Biggr]^{0.55}. 
\quad[{\rm kpc}]
\label{eq:radius}
\end{equation}
Both these relations were derived assuming a Hubble constant $H_0=75$ km/s/Mpc.
Note that one should replace the subscript ``L'' with ``G'',
in equations \ref{eq:be_s79} and
\ref{eq:radius}, when adopting the scenario in which the luminous and dark
components are distributed similarly.

We should stress the uncertainty in using
these empirical $\Omega_{\rm L}$--$M_{\rm L}$--$R_{\rm L}$ relations which are
based upon present-day properties of spheroidals.  Observational constraints
of this ilk, for primeval galaxies, simply do not exist.  For example,
if the proto-galaxy has not fully collapsed at the epoch of galactic winds,
then the binding energy at this point might be a factor of $\sim 2$ less than
that predicted by equation \ref{eq:be_s79} (\cite{AY89});
conversely, mass lost at $t_{\rm GW}$ (either impulsively, or gradually over a
long timescale) would imply that the binding energy at $\sim t_{\rm GW}$ 
might be
anywhere from $\sim 1\rightarrow 4$ times greater than that predicted by
equation \ref{eq:be_s79}.  This latter scenario was explored 
by Hills (1980), Vader (1987), and Angeletti \& Giannone (1991).
We recognise that use of
Saito's (1979a) present-day relationships is not ideal, but to be conservative,
restrict ourselves to them regardless\footnote{\it If \rm the 
$\Omega_{\rm L}$--$M_{\rm L}$--$R_{\rm L}$ relation is really a factor of four
greater than that predicted by equation \ref{eq:be_s79}, then for a
given value of the star efficiency $\nu$, $t_{\rm GW}$ will occur up to an
order of magnitude \it later \rm than normally expected.  This has the
undesired effect of driving the colours/metallicities too red/high; increasing
$\nu$ appropriately to compensate drives $t_{\rm GW}$ down to comparable values
encountered using the canonical expression.  For the analysis which follows,
this difference is not important.}.

\subsection{Supernova remnant interior thermal energy evolution}
\label{ingredsnr}
Another of the primary ingredients to any
SNe-driven galactic wind model is the
assumed evolution of the thermal energy made available to the ISM by each SN
event.  It is the hot, dilute gas in the interior of these SN remnants (SNRs)
which contains virtually all of this thermal energy (\cite{C72}).

For our modeling, we have considered a number of thermal evolutionary scenarios
for SNRs, each of which has its basis in either the classic models of Cox
(1972) and Chevalier (1974) (the \bf ``A'' models\rm, below), 
or the more sophisticated treatment of Cioffi
\etal (1988) (the \bf ``B'' models\rm, below)  The latter models incorporate
additional radiative cooling processes and explicit metallicity effects.  
The relevant equations governing each of the models have already
been outlined in Gibson (1994b,1995).  A cursory treatment can also be found in
Matteucci (1997).  We refer the reader to these papers for the \it minutae\rm,
and for this paper we simply provide the following qualitative listing:

\begin{enumerate}
\item \bf Model A$_0$\rm: SNR shells continue to expand and cool radiatively 
\it ad infinitum \rm (after Cox 1972 and Chevalier 1974).  
The SNR interior thermal energy evolves as
$\varepsilon_{{\rm th}_{\rm SN}}\propto (t/t_{\rm c})^{-0.6}$, where 
$t_{\rm c}$ is the
shell cooling time.  This is the classic scenario adopted by virtually all
previous studies.
\item \bf Model A$_1$\rm: Parallels A$_0$ until the interior pressure in the
SNR is reduced to that of the pressure of the ambient ISM, thereafter merging
and becoming indistinguishable from the surrounding ISM.  No further radiative
cooling of the interior is considered.
\item \bf Model B$_0$\rm: Shells continue to expand and cool radiatively \it ad
infinitum \rm (after Cioffi \etal 1988).
The late-time behaviour of $\varepsilon_{{\rm th}_{\rm SN}}$ varies as
$\propto (t/t_{\rm c})^{-1.0}$.  The additional factor of 
$(t/t_{\rm c})^{-0.4}$
in $\varepsilon_{{\rm th}_{\rm SN}}$ 
is due to Cioffi \etal's (1988) inclusion of
radiative cooling, neglected in the classic Cox (1972) and Chavalier (1974)
studies.  This is the modern analogue to Model A$_0$.
\item \bf Model B$_1$\rm: Shells halt both their expansion and radiative cooling
at the ISM merging time alluded to in Model A$_1$'s description.
\item \bf Model B$_2$\rm: Shells halt their expansion but continue to cool
radiatively \it ad infinitum \rm (\ie 
$\varepsilon_{{\rm th}_{\rm SN}}\propto (t/t_{\rm
c})^{-0.4}$ at later evolutionary stages).
\item \bf Model B$_3$\rm: Parallels Model B$_2$ until reaching a cooling
time beyond which radiative cooling is no longer efficient.
\item \bf Model B$_3^\prime$\rm: Parallels Model B$_3$, unless expanding shells
start coming into contact, and overlapping with, neighbouring shells, in which
case the expansion term is dropped earlier.
\end{enumerate}

The above models can be compared visually by referring to Figure 1 of Gibson
(1994b).
Unless stated otherwise, we shall use Model B$_2$ 
for the late-time
evolution of $\varepsilon_{{\rm th}_{\rm SN}}(t)$.  

%%%%%%%%%%%%%%%%%%%%%%%%%%%%%%%%%%%%%%%%%%%%%%%%%%%%%%%%%%%%%%%%%%%%%%%%%%%%%%%
\section{Analysis}
\label{analysis}

\subsection{The ``template'' models}
\label{template}
In order to examine the sensitivity of model predictions to the various input
ingredients, we first describe a working template of models which, in
general, satisfy the observational constraints outlined in Section
\ref{introduction}.  The pertinent facts are listed in Table
\ref{tbl:templatemodels2}.  We draw attention to the star formation efficiency
$\nu$ in column 2; for the chosen set of input ingredients $\nu$, was treated
as a free parameter and chosen to ensure that the present-day
CML relationships (columns 7 to 10) were recovered, as reflected
by the solid curves in Figures \ref{fig:Zsfr} and \ref{fig:VKsfr}.  This
parallels Arimoto \& Yoshii's (1987) treatment of $\nu$.  Following the
procedure outlined in Gibson (1996a), the V-band
luminosity-weighted metallicity $[<{\rm Z}>]_{\rm V}$ (column 10) is computed
for each model elliptical.

The photometric evolution was coupled to the chemical evolution as outlined in
Gibson (1996a).  In general, the metallicity-dependent isochrones of Worthey
(1994,1995) were adopted, although when those of Bertelli \etal (1994)
were used, the distinction is made.

Our working template was generated using a time- and metallicity-independent 
form
of the Salpeter (1955) IMF, with $m_\ell=0.2$ M$_\odot$ and $m_{\rm U}=65.0$
M$_\odot$.  The metallicity-dependent yields of WW95 were used for Type
II SNe.  The thermal evolution of the ISM was governed by Model B$_2$ (Section
\ref{ingredsnr}).
Following Matteucci (1992), 
diffuse dark matter halos with mass and radial extent ratios
relative to the luminous component of ten were used.

A binary parameter $A=0.03$ 
(equation \ref{eq:Ia_rate}) was chosen \it a posteriori \rm to ensure that the 
Type Ia SN rate (column 11)
was consistent with that observed in the local elliptical
population (\cite{TCB94}).  The stellar population of the giant ellipticals in
our template have [Mg/Fe]$_\ast\approx +0.15$, which is only
marginally lower than
the $\sim +0.2\rightarrow +0.3$ observed (\cite{WFG92}).

Finally, the mass of gas, oxygen, and iron (columns 4, 5, and 6) ejected in the 
galactic wind at $t_{\rm GW}$ (column 3) are also provided in Table
\ref{tbl:templatemodels2}.

\begin{table*}[hpt]
\caption[Template Models: SNe Model B$_2$]{\label{tbl:templatemodels2}
Template Models:  Nucleosynthetic yield
sources are Woosley \& Weaver (1995) for high mass stars, Renzini \& Voli
(1981) for single low and intermediate mass stars, and Thielemann \etal
(1993) for binary-progenitor SNe Type Ia.  Luminosities and colours
derived from Worthey's (1994,1995) isochrones.  SNe remnant thermal energy
follows Model B$_2$ of Section \ref{ingredsnr}, which is derived from Cioffi
\etal (1988).  A single power law initial
mass function, by mass, of slope $x=1.35$, and lower and upper mass limits of
0.2 M$_\odot$ and 65.0 M$_\odot$, respectively, was used.  $\nu$ is the
astration parameter for the star formation rate in equation \ref{eq:sfr}.
The galactic
wind time $t_{\rm GW}$ is in units of Gyrs.  A binary parameter $A=0.03$ was
adopted.  Initial dark-to-luminous mass and radial extents of ten were chosen.
All post-$t_{\rm GW}$ ejecta was retained.
}
\begin{center}
\begin{tabular}{crccccrccrc}
\vspace{2.0mm}
$M_{\rm g}(0)$ & $\nu$$\;\;\;$ & $t_{\rm GW}$ & $m_{\rm g}^{\rm ej}$ &
$m_{\rm O}^{\rm ej}$ & $m_{\rm Fe}^{\rm ej}$ & M$_{\rm V}$ & B-V & V-K & 
$[<{\rm Z}>]_{\rm V}$ & $R_{\rm Ia}$ \\
1.0e6  & 188.9 & 0.006 & 3.5e5  & 2.2e2 & 1.0e1 &  -8.21 & 0.68 & 2.08 & -2.28$\;\;\;\;$ & 0.03 \\
5.0e7  & 209.7 & 0.007 & 1.2e7  & 3.2e4 & 1.5e3 & -12.61 & 0.69 & 2.12 & -1.56$\;\;\;\;$ & 0.03 \\
1.0e9  & 123.1 & 0.016 & 1.7e8  & 3.0e6 & 2.0e5 & -15.89 & 0.74 & 2.44 & -0.51$\;\;\;\;$ & 0.04 \\
5.0e10 &  46.0 & 0.077 & 3.4e9  & 1.1e8 & 9.3e6 & -20.15 & 0.85 & 3.04 & +0.13$\;\;\;\;$ & 0.04 \\
1.0e12 &  17.3 & 0.440 & 1.7e10 & 4.9e8 & 7.5e7 & -23.45 & 0.92 & 3.33 & +0.44$\;\;\;\;$ & 0.05 \\
\end{tabular}
\end{center}
\end{table*}

Figure 1 of Gibson (1996b) illustrates the evolution of elemental abundance of
the primary metals for the $M_{\rm g}(0)=10^{12}$ M$_\odot$ model of Table
\ref{tbl:templatemodels2}.

Figure \ref{fig:1e12_snr} shows the relevant Type II and Ia SNe rates
for this same
$10^{12}$ M$_\odot$ template model.  As expected, beyond $t_{\rm GW}$, the Type
II rate quickly drops to zero as star formation has ceased (and Type II
progenitors have $m\simgt 10$ M$_\odot$), whereas the Type Ia SNe
\footnote{Recall from Section \ref{progenitors} that in the binary Type Ia SNe
scenario, ratios of secondary-to-primary mass of $\sim 1$ are favoured, and
that the explosion ``clock'' is set by the lifetime of the secondary.  Imposing
the chosen
3 M$_\odot$ minimum-system-mass means that beyond $\sim$3 Gyr (\ie the
lifetime of a 1.5 M$_\odot$ star), one becomes restricted to mass ratios less
than unity, with the ratio getting smaller and smaller as time goes by, and
thus the distribution function $f(\mu)$ of equation \ref{eq:Ia_rate} favours
less probable values.  This results in an expected ``break'' and subsequent
``downturn'' in the Type Ia SNe rate at precisely the point seen in Figure
\ref{fig:1e12_snr}.  This downturn is also seen in Figure 1 of Greggio \&
Renzini (1993).}
continue right up until the present-day (the value shown at $t=12$ Gyr
reflecting the observed value in local ellipticals 
-- \cite{TCB94}).  
The first Type Ia SNe do not appear until $t\approx 35$
Myr (lifetime of an 8 M$_\odot$ star -- the most massive secondary allowed in
our SNe Ia formalism -- Section \ref{progenitors}), hence the delay in $R_{\rm
Ia}$, relative to $R_{\rm II}$, seen in Figure \ref{fig:1e12_snr}.

\begin{figure}[ht]
\centering
\vspace{8.5cm}
\includegraphics{`cat}
\caption[]{Time-dependence of the Type Ia and II SNe
rates for the $M_{\rm g}(0)=10^{12}$
M$_\odot$, SNe energy Model B$_2$, template.  The galactic wind epoch $t_{\rm
GW}$ (and hence cessation of star formation) is noted.
\label{fig:1e12_snr}}
\end{figure}

\subsection{Influence of the ingredients}
\label{influence}

The following six subsections provide at least a cursory examination of the
influence of each of the major input ingredients.  As the parameter space can
be large, we will restrict the analysis to just a few pertinent examples, from
which the importance of the relevant parameter can be addressed.  Of particular
importance
will be the effect upon $t_{\rm GW}$, as this sets the temporal extent of
the star formation phase in these galactic models, and thus influences heavily
the predicted present-day photo-chemical properties of the resultant stellar
populations.  We shall see that the results are sensitive to some of the
ingredients, and insensitive to others.

\subsubsection{Stellar lifetimes}
\label{analtau}

Using the least and most massive template models of Table
\ref{tbl:templatemodels2}, we now vary the stellar lifetime formalism in order
to view its influence upon the predicted
present-day photometric and chemical properties.  Four different forms were
tried -- the two Schaller \etal (1992) forms (Z=0.001 and Z=0.020:
SSMM92$_1$ and SSMM92$_2$, respectively); 
G\"usten \& Mezger (1982, hereafter GM82); and Talbot \& Arnett (1971, 
hereafter TA71).  Each were shown in Figure \ref{fig:tau}.
Re-examination of this figure should lead one to the intuitive conclusion that
the results, for all but perhaps the TA71 ``single power-law'',
should be relatively insensitive to the chosen $\tau$ form. 

Table \ref{tbl:tau_tbl} shows in a more quantitative sense that this is true.
For discussion purposes, and the sake of brevity, let us
restrict ourselves to the $M_{\rm g}(0)=10^{12}$ M$_\odot$ model.  Very little
difference is encountered when switching between the GM82 and
SSMM92$_{1,2}$ lifetimes -- recall from Section \ref{ingredtau}
that the former was
based upon older, super-solar metallicity (Z=0.03),
stellar models, whereas the
latter allows for a more modern prescription for stellar evolution theory, and
two different metallicities (Z=0.001 and Z=0.020).  The lower metallicity
SSMM92$_1$ form leads to marginally later wind epochs ($\sim 5$\% later), 
which in turn leads
to marginally more iron being ejected ($\sim 8$\% more), due to the increased
importance of Type Ia SNe at $t\simgt 0.3$ Gyr.  The stellar lifetimes of Type
II SNe progenitors for the Z=0.001 SSMM92 models are slightly longer
than those for solar metallicity, by typically $\sim 10$\%.  The main reason
for this is that the lower metallicity models burn at a reduced luminosity,
thereby lengthening their main sequence lifetimes.

\begin{table*}[hpt]
\caption[Stellar Lifetime Sensitivity]{\label{tbl:tau_tbl}
Influence of stellar lifetime selection upon the galactic wind time $t_{\rm
GW}$ (and hence, gas, O, and Fe, ejection, as well as resultant photometric
properties).  $\tau$ denotes the lifetime formalism adopted (all other
ingredients as in Section \ref{template}.  TA71=\cite{TA71};
GM82=\cite{GM82}; SSMM92=\cite{SSMM92} -- subscripts `1' and `2' refer
to Z=0.001 and Z=0.020 lifetimes, respectively.  Superscript $\ast$ implies
that $\nu$ has been adjusted from that shown in Table \ref{tbl:templatemodels2}
in order to recover the same present-day photo-chemical properties shown in the
earlier template table.
}
\begin{center}
\begin{tabular}{lcccccrccr}
\vspace{2.0mm}
$M_{\rm g}(0)$ & $\tau$ & $t_{\rm GW}$ & $m_{\rm g}^{\rm ej}$ &
$m_{\rm O}^{\rm ej}$ & $m_{\rm Fe}^{\rm ej}$ & M$_{\rm V}$ & B-V & V-K & 
$[<{\rm Z}>]_{\rm V}$ \\
1.0e6  &  SSMM92$_2$ & 0.006 & 3.5e5  & 2.2e2 & 1.0e1 &  -8.21 & 0.68 & 2.08 & -2.28$\;\;\;\;$ \\
1.0e6  &  SSMM92$_1$ & 0.006 & 3.2e5  & 1.5e2 & 5.1e0 &  -8.25 & 0.68 & 2.08 & -2.49$\;\;\;\;$ \\
1.0e6$^\ast$  &  SSMM92$_1$ & 0.006 & 3.4e5  & 2.3e2 & 1.0e1 &  -8.22 & 0.68 & 2.08 & -2.27$\;\;\;\;$ \\
1.0e12 &  SSMM92$_2$ & 0.440 & 1.7e10 & 4.9e8 & 7.5e7 & -23.45 & 0.92 & 3.33 & +0.44$\;\;\;\;$ \\
1.0e12 &  SSMM92$_1$ & 0.462 & 1.7e10 & 4.9e8 & 8.1e7 & -23.47 & 0.92 & 3.33 & +0.44$\;\;\;\;$ \\
1.0e12 &  GM82       & 0.454 & 1.7e10 & 5.0e8 & 8.2e7 & -23.47 & 0.93 & 3.35 & +0.44$\;\;\;\;$ \\
1.0e12 &  TA71       & 0.345 & 2.1e10 & 6.5e8 & 7.1e7 & -23.44 & 0.86 & 3.11 & +0.35$\;\;\;\;$ \\
1.0e12$^\ast$ &  TA71& 1.476 & 1.1e10 & 2.5e8 & 4.6e7 & -23.47 & 0.92 & 3.34 & +0.48$\;\;\;\;$ \\
\end{tabular}
\end{center}
\end{table*}

One might be tempted to infer that the galactic wind epoch should be earlier
for the higher metallicity $\tau$
assumption because of the shorter stellar lifetimes,
and therefore, earlier SN explosion.  As can be seen in Table
\ref{tbl:tau_tbl}, this is indeed the case.  

Not surprisingly, the extremely different form for the TA71
stellar lifetimes, already encountered in Figure \ref{fig:tau}, leads to more
substantial differences in the output models.  The second-to-last entry in
Table \ref{tbl:tau_tbl} shows that for $\nu=17.3$ Gyr$^{-1}$, the TA71
$\tau$-formalism leads to earlier galactic wind times ($\sim 25$\% earlier)
than that found with the SSMM92 one, regardless of metallicity.
In fact, the wind occurs early enough for the chosen model parameters, that the
metallicity evolution does not continue to a late-enough epoch to ensure that
the final stellar population's colours are consistent with those seen today --
\eg the final V-K is $\sim 0.25$ mag too blue.

We might naively have expected 
the opposite behaviour, as the TA71 lifetimes
range anywhere from 0 to $\sim$6 
times longer than those found by SSMM92,
and thus the typical Type II SN explosion would occur later, leading one to
perhaps expect a later wind.  This is not what
is encountered, the primary reason being that the substantially longer
lifetimes encountered using 
TA71 means that the bulk of the Type II SNe explosions
are delayed by many tens of millions of years in comparison with the
SSMM92 lifetimes.  This delay means that the bulk of the
``TA71'' explosions occur in a gaseous medium whose density has been
depleted by the ongoing star formation during the ``delay''.  A reduced density
leads to a greater SNR cooling time, and thus
a greater thermal energy contribution per SN event, which in turn leads to an
earlier wind time.

The final entry in Table \ref{tbl:tau_tbl} shows the implications of \it
forcing \rm the $M_{\rm g}(0)=10^{12}$ M$_\odot$ model to coincide with that
seen locally in giant ellipticals.  This was done by fixing all the parameters,
save the star formation efficiency $\nu$, which was reduced from 17.3
Gyr$^{-1}$ to 7.7 Gyr$^{-1}$.  This leads to a wind epoch which is three to
four times later, but it does mean that the mean photo-chemical properties of
the stellar populations have enough time to evolve to that seen locally.
Because the ejection phase occurs so late, it should not be surprising to note
that even with the reduced star formation efficiency, the total gas mass
ejected is down $\sim 35$\%, and the ejecta's [O/Fe] is also reduced by $\sim
0.15$
dex (again, due to the increased contribution from the iron-important,
longer-lived, Type Ia SNe), in comparison with the SSMM92$_2$ model.

The choice of
stellar lifetime does not influence the resultant photo-chemical predictions
by more than a few percent (provided the discrepant singular power-law form is
avoided).  Future versions of {\bf MEGaW} will incorporate a
fully self-consistent metallicity-dependent lifetimes, a l\`a Bazan \& Mathews
(1990), but
for the problems at hand, this does not appear to be a pressing need.

\subsubsection{Stellar nucleosynthesis yields}
\label{analyields}

Let us now turn our attention to one of the primary input ingredients -- the
Type II SNe nucleosynthesis yields.  The basic data were introduced in Section
\ref{ingredyields}, to which the reader is referred for specifics.  Table
\ref{tbl:tbl_yields} shows how the galactic wind epoch changes as a function of
yield selection for the five sources.  We only show the results for one initial
mass -- $M_{\rm g}(0)=10^{12}$ M$_\odot$.

The star formation efficiency parameter $\nu=17.3$ Gyr$^{-1}$ for all entries
in the table without an asterisk in the first column.  Recall that this
was the value required to ensure that the $10^{12}$ M$_\odot$ model, 
in conjunction
with the WW95 yields and the other parameters in the template models
(Section \ref{template}), ended up with V-K$\approx 3.35$ and $[<{\rm
Z}>]_{\rm V}\approx +0.4$ by the present-day.  This is reflected by
the first entry to Table \ref{tbl:tbl_yields}.

\begin{table*}[hpt]
\caption[Yields Sensitivity]{\label{tbl:tbl_yields}
Influence of stellar nucleosynthesis yields
selection upon the galactic wind time $t_{\rm
GW}$ (and hence, gas, O, and Fe, ejection, as well as resultant photometric
properties).  $m_{\rm Z}^{\rm ej}$ 
denotes the Type II SNe yields source (all other
ingredients as in Section \ref{template}.  A91=\cite{A91}; LH95=\cite{LH95};
WW95=\cite{WW95}; TNH95=\cite{TNH95}; M92=\cite{Md92}.
Superscript $\ast$ implies
that $\nu$ has been adjusted from that shown in Table \ref{tbl:templatemodels2}
in order to recover the same present-day photo-chemical properties shown in the
earlier template table.
}
\begin{center}
\begin{tabular}{lcccccrccr}
\vspace{2.0mm}
$M_{\rm g}(0)$ & $m_{\rm Z}^{\rm ej}$ & $t_{\rm GW}$ & $m_{\rm g}^{\rm ej}$ &
$m_{\rm O}^{\rm ej}$ & $m_{\rm Fe}^{\rm ej}$ & M$_{\rm V}$ & B-V & V-K & 
$[<{\rm Z}>]_{\rm V}$ \\
1.0e12 &  WW95   & 0.440 & 1.7e10 & 4.9e8 & 7.5e7 & -23.45 & 0.92 & 3.33 & +0.44$\;\;\;\;$ \\
1.0e12 &  TNH95  & 0.463 & 1.6e10 & 5.5e8 & 7.4e7 & -23.38 & 0.94 & 3.48 & +0.49$\;\;\;\;$ \\
1.0e12$^\ast$ &  TNH95  & 0.161 & 2.5e10 & 9.7e8 & 8.9e7 & -23.36 & 0.91 & 3.35 & +0.42$\;\;\;\;$ \\
1.0e12 &  A91    & 0.460 & 1.6e10 & 4.6e8 & 7.3e7 & -23.39 & 0.94 & 3.47 & +0.48$\;\;\;\;$ \\
1.0e12$^\ast$ &  A91    & 0.160 & 2.5e10 & 8.0e8 & 8.2e7 & -23.36 & 0.91 & 3.35 & +0.40$\;\;\;\;$ \\
1.0e12 &  LH95   & 0.406 & 2.0e10 & 2.3e8 &  n/a  & -23.45 & 0.93 & 3.29 & +0.28$\;\;\;\;$ \\
1.0e12$^\ast$ &  LH95   & 0.736 & 1.5e10 & 1.9e8 &  n/a  & -23.48 & 0.94 & 3.33 & +0.30$\;\;\;\;$ \\
1.0e12 &  M92    & 0.454 & 1.6e10 & 2.8e8 &  n/a  & -23.40 & 0.96 & 3.45 & +0.43$\;\;\;\;$ \\
1.0e12$^\ast$ &  M92    & 0.159 & 2.6e10 & 4.4e8 &  n/a  & -23.37 & 0.92 & 3.33 & +0.39$\;\;\;\;$ \\
\end{tabular}
\end{center}
\end{table*}

One thing we note immediately from Table \ref{tbl:tbl_yields} is that for
a given $\nu=17.3$ Gyr$^{-1}$, $t_{\rm
GW}$ is not particularly
sensitive to yield compilation, except that using LH95
leads to wind times which are $\sim 10$\% earlier than the other four.
This can be understood by recognising that over a wide range of Type II SNe
progenitor masses, the LH95 yields are typically $\sim 30\rightarrow
40$\% lower than the others.
This is due to the combination of their inclusion
of stellar winds, and their semi-convection treatment (Ledoux criterion and 
minimal overshooting -- Gibson 1997, in preparation).

The top panel of Figure \ref{fig:sal_nuc_nu17} illustrates how the reduced Z
yield favoured by LH95 manifests itself in the ISM evolution -- Z$_{\rm
g}$ is consistently $\sim 30\rightarrow 40$\% lower.  The
cooling time of a SNR shell increases as a function of decreasing 
metallicity (Cioffi \etal 1988; Gibson 1995), and thus
the lower ISM Z$_{\rm g}$ encountered
with LH95 means that radiative cooling of each SN event is
delayed relative to the other yield compilations.  This greater energy per SN
is what is responsible for the slightly earlier $t_{\rm GW}$.
We note in passing that the earlier $t_{\rm GW}$ means that the mass of gas
ejected in the galactic wind is $\sim 20$\% greater for a given $\nu$.

\begin{figure}[ht]
\centering
\vspace{10.0cm}
\includegraphics{`cat}
\caption[]{Evolution of the ISM metallicity 
Z$_{\rm g}$, and C/O and O/Fe ratios as a function of
time for $\nu=17.3$ Gyr$^{-1}$.  This $\nu$ value ensures the ``Woosley \&
Weaver (1995)''
model's present-day photo-chemical properties are consistent with those 
observed locally.  Evolution ceases at $t_{\rm GW}$.
\label{fig:sal_nuc_nu17}}
\end{figure}

Not surprisingly, the LH95 $\nu=17.3$ Gyr$^{-1}$ oxygen ejected at $t_{\rm GW}$
is $\sim 55$\% lower than that found using the WW95, TNH95, or
A91 yields.  A similar lower oxygen ejecta mass is found with the M92 yields.
Both LH95 and M92
have reduced oxygen yields due to stellar
winds, and this is reflected in the oxygen ejected at $t_{\rm GW}$ in Table
\ref{tbl:tbl_yields}.  Note though that while both LH95 and
M92 models have lower oxygen ejected at $t_{\rm GW}$, only LH95's has
a similarly reduced ``Z'' yield.  M92's Z yield is boosted by a greatly
enhanced carbon contribution, especially for solar metallicity and
masses $m\simgt 25$ M$_\odot$.

Again, just restricting ourselves to the $\nu=17.3$ Gyr$^{-1}$ numbers for the
time being, we see that 
the TNH95 oxygen ejecta appears to be $\sim 10$\% greater than the other models
run without stellar winds (\ie A91 and WW95).  This may or may
not be so -- an uncertainty is introduced when using the TNH95
yields because the most massive model in their compilation is only 25
M$_\odot$.  We have just extrapolated beyond this last point in order to
estimate the most massive Type II SNe yields (Section
\ref{ingredyields}), which may not be optimal.  Note
that the TNH95
oxygen yields closely parallel WW95 up to $m=25$ M$_\odot$;
anything beyond that is at best a rough estimate.  This problem 
will be ever-present with the TNH95 yields due to its limited mass coverage.

The bottom panel for Figure \ref{fig:sal_nuc_nu17} shows that the ISM [O/Fe]
converges toward $\sim +0.0\rightarrow +0.1$ beyond $t\approx 0.3$ Gyr,
regardless of yield compilation.  The TNH95 curve is $\sim 0.1$ dex
greater than the A91 and WW95 at $t_{\rm GW}$, and considerably
higher for $t\simlt 0.03$ Gyr.  Again, this is due primarily to the uncertain
extrapolation to $m>25$ M$_\odot$ -- the TNH95 iron yield extrapolates
to very small values, but more importantly, the oxygen
yield grows very large beyond $m\approx 35$ M$_\odot$.
This latter extrapolation tends to drive C/O in the earliest phases of the
evolution to very low values (see the middle panel of Figure
\ref{fig:sal_nuc_nu17}).

This middle panel of Figure \ref{fig:sal_nuc_nu17}
is a nice illustration of how the uncertainties in the stellar
evolution can manifest itself.  Whereas we saw that [O/Fe] only spans $\sim
+0.1\rightarrow +0.3$ dex for the three yield compilations with iron for
$t\simgt 0.1$ Gyr, the C/O evolution is far more sensitive to the exact
compilation used.  C/O is seen to span almost a full dex during
the same time period.  Unlike the others, the WW95 curve has an initial
decline in the C/O evolution due to the high mass Z=0.0000 stars which have
very high C/O.  The upturn in C/O beyond $t\approx
0.1$ Gyr is due primarily to the increased importance of
intermediate mass stars undergoing multiple dredge-ups with stellar winds and
PNe ejection.  The enormous metallicity-sensitivity in the high
mass carbon yields of M92 leads to the 
steeper slope for times $t\simlt 0.1$ Gyr, when compared with the other four
yield sources.

Only the LH95 yields required a lower value for $\nu$, albeit the
adjustment was minor, the colours being reddened by only a few hundredths
of a magnitude in V-K.  The other three (besides the template WW95)
required a factor of $\sim 2$
increase in the star formation efficiency, the result of which was a galactic
wind occurring $\sim 0.3$ Gyr earlier.  This had the desired effect of
``bluing'' the predicted V-K by $\sim 0.15$ mag, in better agreement with the
mean of the observations shown in Figure \ref{fig:VKsfr}.

\subsubsection{Thermal and binding energy evolution of the ISM}
\label{analenergy}

There are a number of inter-related aspects to the ISM energetics 
(thermal+gravitational binding) which have
already been touched upon in Section \ref{ingredenergy}.  We will now look at a
few examples which illustrate the sensitivity of $t_{\rm GW}$ to the assumed
dark matter distribution (diffuse halo (DH) or dynamically dominant (DM)), as
well as a
couple of ``hidden'' uncertainties that plague many galactic wind models,
but which are not widely appreciated.  Arguments pertaining to the influence of
assumed individual SNR energetics are delayed until Section
\ref{analsnr}.

Table \ref{tbl:dh} shows how $t_{\rm GW}$, and the relevant present-day
photo-chemical properties, vary with assumed initial dark-to-luminous mass
and radial extent ratios.  Recall that $M_{\rm D}/M_{\rm L}\equiv R_{\rm
D}/R_{\rm L}\equiv 10$ for our template models.  It is readily apparent that
the results are not sensitive to the dark matter content, provided it is
distributed diffusely, following the prescription of Bertin \etal (1992).
This is not
surprising given the numerical example of Section \ref{ingredenergy}, 
which showed the dominance of the luminous-luminous interaction term, as
well as the
results of Matteucci (1992).  Obviously, for models in which the dark matter is
more centrally condensed (\eg D=10; R=3), the potential is deeper
(equation \ref{eq:be_BSS}), and
it consequently takes longer for the ISM
thermal energy to build up to the
necessary level to overcome the binding energy, but the difference is not
extreme, and could be minimised with a marginal increase in $\nu$.

\begin{table*}[hpt]
\caption[Model Sensitivity to Diffuse Dark Halo Distributions]
{\label{tbl:dh} Influence of diffuse dark matter halo distribution 
upon the galactic wind time $t_{\rm
GW}$ (and hence, gas, O, and Fe, ejection, as well as resultant photometric
properties), for the $M_{\rm g}(0)=10^{12}$ M$_\odot$ model.  
D$\equiv M_{\rm D}/M_{\rm L}$; R$\equiv R_{\rm D}/R_{\rm L}$.
All other ingredients as described in Section \ref{template}.
}
\begin{center}
\begin{tabular}{ccrccccrcccc}
\vspace{2.0mm}
D & R & $\nu$$\;\;\;$ & $t_{\rm GW}$ & $m_{\rm g}^{\rm ej}$ &
$m_{\rm O}^{\rm ej}$ & $m_{\rm Fe}^{\rm ej}$ & M$_{\rm V}$ & B-V & V-K & 
$[<{\rm Z}>]_{\rm V}$ & $<[{\rm Mg/Fe}]>_{\rm V}$ \\
10        & 10 &  17.3 & 0.440 & 1.7e10 & 4.9e8 & 7.5e7 & -23.45 & 0.92 & 3.33 & +0.44 & +0.13 \\
10        &  3 &  17.3 & 0.719 & 8.5e9  & 2.1e8 & 4.4e7 & -23.49 & 0.93 & 3.37 & +0.49 & +0.08 \\
 3        & 10 &  17.3 & 0.382 & 2.2e10 & 6.6e8 & 9.2e7 & -23.44 & 0.92 & 3.32 & +0.41 & +0.13 \\
 0        & n/a&  17.3 & 0.357 & 2.5e10 & 7.7e8 & 1.0e8 & -23.43 & 0.92 & 3.31 & +0.39 & +0.13 \\
\end{tabular}
\end{center}
\end{table*}

The situation is not particularly different
if one makes the assumption that the dark matter
is distributed similarly to the luminous matter, our so-called dynamically
dominant (DD) model.  In this case the gaseous binding energy follows equation
\ref{eq:gasbe} (after \cite{S79b}).  Table \ref{tbl:dd} shows how the model
predictions for the template $10^{12}$ M$_\odot$ model vary for initial
dark-to-luminous mass ratios D=0, 3, and 10 for the DD model.

\begin{table*}[hpt]
\caption[Model Sensitivity to Dynamically Dominant Dark Matter Distributions]
{\label{tbl:dd} Influence of dark matter, distributed similarly to the luminous
component, upon the galactic wind time $t_{\rm
GW}$ (and hence, gas, O, and Fe, ejection, as well as resultant photometric
properties), for the $M_{\rm g}(0)=10^{12}$ M$_\odot$ model.  
D$\equiv M_{\rm D}/M_{\rm L}$.
All other ingredients as described in Section \ref{template}.
}
\begin{center}
\begin{tabular}{crccccrcccc}
\vspace{2.0mm}
D & $\nu$$\;\;\;$ & $t_{\rm GW}$ & $m_{\rm g}^{\rm ej}$ &
$m_{\rm O}^{\rm ej}$ & $m_{\rm Fe}^{\rm ej}$ & M$_{\rm V}$ & B-V & V-K & 
$[<{\rm Z}>]_{\rm V}$ & $<[{\rm Mg/Fe}]>_{\rm V}$ \\
 0        &  17.3 & 0.569 & 1.2e10 & 3.0e8 & 5.6e7 & -23.47 & 0.93 & 3.35 & +0.47 & +0.11 \\
 3        &  17.3 & 0.453 & 1.6e10 & 4.6e8 & 7.2e7 & -23.45 & 0.92 & 3.34 & +0.44 & +0.13 \\
10        &  17.3 & 0.421 & 1.9e10 & 5.3e8 & 8.0e7 & -23.45 & 0.92 & 3.33 & +0.43 & +0.13 \\
\end{tabular}
\end{center}
\end{table*}

It can be seen that
$t_{\rm GW}$ is not unduly influenced by the presence of a DD dark matter
component.  This is at odds with the conclusion of Matteucci (1992)
who found very
late wind epochs ($t_{\rm GW}\simgt 9$ Gyr) for $M_{\rm g}(0)=10^{12}$
M$_\odot$, with D=10 (her Model C2).  The source of the confusion can be traced
to equation \ref{eq:gasbe} --  when the dark matter was
distributed similarly to the luminous matter, Matteucci had $M_{\rm g}(0)$ 
incorrectly in place of $M_{\rm G}$ in the denominator \it outside \rm the 
brackets.  The proper form is as shown in equation \ref{eq:gasbe} (and was also
laid down in its proper form by \cite{FP93}, following their equation 1).
By using the luminous mass,
as opposed to the gravitational mass, this leads to an order of magnitude
overestimation of the
gaseous binding energy (through equation \ref{eq:gasbe}).

In fact, as inspection of column 3 of Table \ref{tbl:dd} shows, $t_{\rm GW}$
actually decreases marginally as one goes to higher dark matter contents,
simply because the corresponding increase in SNe energy efficiency 
\footnote{The increase in SN energy deposition efficiency with increasing total
mass, simply comes about because in order to honour equation \ref{eq:radius},
as $M_{\rm G}$ increases, for a given $M_{\rm L}$, there must also be a
corresponding increase in $R_{\rm G}$.  By spreading the same $M_{\rm L}$ over
a larger volume of radius $R_{\rm G}$, the resulting decrease in the hydrogen
number density
means a corresponding increase in both the SN shell cooling time and the ISM
merging time,
through equations 9 and 14, respectively, of Gibson (1994b),
and thus an increase in $\varepsilon_{{\rm th}_{\rm SN}}$, through
equation 12 of Gibson 1994b.}  
outweighs the accompanying increase in the
gaseous binding energy (through equation \ref{eq:gasbe}).
This behaviour, while perhaps interesting on a
``mathematical'' level, may or may not be wholly relevant on a ``physical''
level.  Specifically, at some point, the validity of
continually increasing R$_{\rm G}$ with M$_{\rm G}$, for a given M$_{\rm L}$,
must be called into question, as the predicted surface brightnesses will be at
odds with the observations.  On top of this, as stressed by Bertin \etal
(1994), and references therein, the dynamics of elliptical galaxies are better
explained by assuming the presence of a massive dark halo.  For all of these
reasons, we will henceforth be restricting ourselves to the ``dark halo''
formalism of Bertin \etal (1992), in order to model the evolution of the
gaseous binding energy.

One last ``hidden'' factor which can alter the calculated value of
$t_{\rm GW}$, is the formulation used for estimating the hydrogen number
density $n_0$ -- a prime ingredient in estimating the amount of energy per SN
event which is made available to the ISM for driving a galactic wind, via its
role in setting the cooling time for an individual SNR, 
$n_0$ should be based upon the ISM density
at the time of the SN explosion (\ie $n_0$ an explicit function of time).  In
order to simplify the energy calculations some early models 
(\eg \cite{AY87}; \cite{MT87}) used $n_0(t)\equiv
n_0(0)$, for all time $t\ge 0$.  This point was first alluded to by Angeletti \&
Giannone (1990), and more recently by Gibson (1996b).  The latter reference
quantifies its effect as pertaining to Arimoto \& Yoshii's (1987) models, and
the reader is directed there for more details.

\subsubsection{Supernova remnant interior thermal energy evolution}
\label{analsnr}

One of the more interesting aspects of our work to date has been the
re-examination of the role played by supernova remnant (SNR) thermal energy in
powering galactic winds.  As already alluded to in Section \ref{ingredsnr},
the older galactic wind models (\eg \cite{AY87};
\cite{MT87}; \cite{BCF94})
adopt the classic Cox (1972) and Chevalier (1974)
formalism for the energetics (our so-called ``A'' models).  
This form has since been
supplanted by the more sophisticated models of Cioffi \etal (1988)
(what we term the ``B'' models), and our work is
the first to incorporate these improvements.  
Not only is the evolution of the individual SNR 
affected by this new formalism, the influence of overlapping shells has at
least been treated to first-order.  This turns out to be a 
crucial point, which was
recognised in Larson's (1974b) 
seminal paper, but again, not fully appreciated in
many of the subsequent detailed models, 
save those of Dekel \& Silk (1986), Babul \& Rees (1992), and
Nath \& Chiba (1995).  Gibson (1994b,1995) has already examined some of the
implications of the new formulations, and as such, only a few important new
points will be made here.

The adopted energetics form can play a
vital role in setting the galactic wind time (and consequently the
end of star formation, and the resultant photo-chemical properties).
Table \ref{tbl:snr_tbl} illustrates how sensitive $t_{\rm GW}$ is to the
SNR thermal energy evolution model.  For brevity, we restrict ourselvesto an
analysis of the $M_{\rm g}(0)=10^{12}$ M$_\odot$ model.

Recalling that Model B$_2$ has been adopted for the template models (Section
\ref{template}), we see that its entry in Table \ref{tbl:snr_tbl} yields
$t_{\rm GW}=0.44$ Gyr, with the appropriate present-day photo-chemical
properties.  Model A$_0$ is virtually indistinguishable from B$_2$
for this example, which is not surprising when a graphical comparison of the
different $\varepsilon_{{\rm th}_{\rm SN}}$ models is examined (see Figure 1 of
\cite{G94b}).  Model B$_0$ (Cioffi \etal's 1988 direct analog to the classic
Model A$_0$, but including radiative cooling of the interior and metallicity
effects) leads to very late galactic wind times ($t_{\rm GW}\approx
7\rightarrow 8$ Gyr, for giant ellipticals), which for the model presented here
would imply star formation rates of $\psi\approx 15$
M$_\odot$/yr at redshifts $z\sim 0.4$, apparently at odds with the observations
(\cite{S86}).

Models A$_1$ and B$_1$ lead to significantly earlier wind times as the
late-time evolution of each individual remnant differs from the continual
energy-loss models A$_0$ and B$_2$.  
For the same $\nu$, this of course results in
colours/metallicities which are too blue/low in comparison with the template
Model B$_2$.  Model B$_1^\ast$ illustrates the results when $\nu$ is reduced to
2.9 Gyr$^{-1}$, in order to recover the proper photo-chemical properties.  The
later $t_{\rm GW}$ helps these properties, but at the expense, somewhat, of the
predicted stellar [Mg/Fe].  More importantly, the reduced $\nu$ leads to an
order of magnitude more gas being ejected at $t_{\rm GW}$, despite its later
occurrence.  

\begin{table*}[hpt]
\caption[$M_{\rm g}(0)=10^{12}$ M$_\odot$ Model Sensitivity to SNR Thermal
Energy] {\label{tbl:snr_tbl} Influence of SNR thermal energy scenario
upon the galactic wind time $t_{\rm
GW}$ (and hence, gas, O, and Fe, ejection, as well as resultant photometric
properties), for the $M_{\rm g}(0)=10^{12}$ M$_\odot$ model
-- ``A'' models are based upon Cox (1972) and Chevalier (1974); 
``B'' models are based upon Cioffi \etal (1988).  See text for details.
All other ingredients as described in Section \ref{template}.
Superscript $\ast$ implies
that $\nu$ has been adjusted from that shown in Table \ref{tbl:templatemodels2}
in order to recover the same present-day photo-chemical properties shown in the
earlier template table (specifically, (V-K) and $[<{\rm Z}>]_{\rm V}$).
}
\begin{center}
\begin{tabular}{crccccrcccc}
\vspace{2.0mm}
Model & $\nu$$\;\;\;$ & $t_{\rm GW}$ & $m_{\rm g}^{\rm ej}$ &
$m_{\rm O}^{\rm ej}$ & $m_{\rm Fe}^{\rm ej}$ & M$_{\rm V}$ & B-V & V-K & 
$[<{\rm Z}>]_{\rm V}$ & $<[{\rm Mg/Fe}]>_{\rm V}$ \\
A$_0$      &  17.3 & 0.453 & 1.6e10 & 4.6e8 & 7.2e7 & -23.45 & 0.92 & 3.34 & +0.44 & +0.13 \\
A$_1$      &  17.3 & 0.115 & 2.0e11 & 4.2e9 & 3.4e8 & -23.22 & 0.87 & 3.03 & $\,\,$-0.09 & +0.10 \\
B$_0$      &  17.3 & 7.355 & 6.5e8  & 1.5e7 & 3.5e6 & -23.55 & 0.94 & 3.45 & +0.45 & $\,\,$-0.16 \\
B$_1$      &  17.3 & 0.151 & 1.3e11 & 3.4e9 & 2.8e8 & -23.30 & 0.89 & 3.13 & +0.05 & +0.10 \\
B$_1^\ast$ &   2.9 & 0.922 & 1.6e11 & 3.9e9 & 4.1e8 & -23.32 & 0.94 & 3.31 & +0.12 & +0.04 \\
B$_2^\ast$ &  17.3 & 0.440 & 1.7e10 & 4.9e8 & 7.5e7 & -23.45 & 0.92 & 3.33 & +0.44 & +0.13 \\
B$_3$      &  17.3 & 0.270 & 4.5e10 & 1.4e9 & 1.5e8 & -23.40 & 0.91 & 3.28 & +0.30 & +0.13 \\
B$_3^\prime$& 17.3 & 0.206 & 7.7e10 & 2.3e9 & 2.1e8 & -23.36 & 0.90 & 3.22 & +0.19 & +0.12 \\
B${_3^\prime}^\ast$& 7.7 & 0.539 & 6.4e10 & 1.9e9 & 2.1e8 & -23.39 & 0.94 & 3.35 & +0.27 & +0.10 \\
\end{tabular}
\end{center}
\end{table*}

Model B$_3$ is perhaps the best representation of an \it individual \rm
SNR's evolution
(\cite{G94b}), and the predicted wind time and properties
are intermediate to the extreme models A$_1$ and B$_1$, and the template B$_2$.
Of course, SNRs do not evolve in isolation; they eventually come into contact
with neighbouring shells, overlap, and subsequent SNe explosions can occur in
the subsequent rarefied bubble (or superbubble -- \eg \cite{T92}).  Model
B$_3^\prime$ takes into account, roughly, the shell overlap effects.
Similarly, Model
B${_3^\prime}^\ast$ uses a reduced $\nu=7.7$ Gyr$^{-1}$ to better reproduce the
photo-chemical properties.  This model, as well as the Models B$_1^\ast$ and
B$_2^\ast$ illustrate that treatment of the individual SNR energetics is an
integral component in predicting the mass of gas ejected at $t_{\rm GW}$.
Ellipticals with similar final photo-chemical properties can differ in this
prediction by up to an order of magnitude.  

\subsubsection{Initial mass function}
\label{analimf}

We can anticipate the influence of IMF selection by simply
looking at the mass fraction tied up in Type II SNe progenitors (\ie $m\simgt
12$ M$_\odot$), as, for example, these are the primary source of
$\alpha$-elements in chemical evolution models, as well as the major
contributor to the ISM energetics.  Table
\ref{tbl:imf_frac} lists the mass fraction of stars $m\ge 12$ M$_\odot$ for
each of the four primary IMFs
under consideration, assuming a total mass range of $0.2\rightarrow 65.0$
M$_\odot$.  One can see that an order of magnitude exists between the extrema,
with the flatter Arimoto \& Yoshii (1987, hereafter AY87)
IMF having $\sim 3$, $\sim 5$, and $\sim 10$ times the mass of Salpeter's
(1955, hereafter S55), Kroupa \etal's (1993, hereafter KTG93), and 
Scalo's (1986, hereafter S86) IMFs, respectively,
locked into Type II SNe progenitors of initial mass $m\ge 12$ M$_\odot$.

\begin{table}[t]
\caption[IMF Massive Star Fractions]{\label{tbl:imf_frac}
The mass fraction $f_m$
tied up in stars more massive than 12 M$_\odot$ in the four
primary IMFs under consideration in {\bf MEGaW}.  A mass range of 0.2 to 65.0
M$_\odot$ was assumed.  S55=Salpeter (1955); AY87=Arimoto \& Yoshii (1987);
KTG93=Kroupa, Tout \& Gilmore (1993); S86=Scalo (1986).}
\begin{center}
\begin{tabular}{cc}
\vspace{2.0mm}
IMF & $f_m(m\ge 12 {\rm M}_\odot)$ \\
AY87  & 0.323 \\
S55   & 0.123 \\
KTG93 & 0.061 \\
S86   & 0.035 \\
\end{tabular}
\end{center}
\end{table}

In Table \ref{tbl:imf_tbl} we start with the template $10^{12}$ M$_\odot$ model
(from Section \ref{template}), but vary in turn the IMF selection 
between the four listed in Table \ref{tbl:imf_frac}.
Parallel sets of models were run for two different
SNe thermal energy forms -- Models A$_0$ and B$_2$ of Section
\ref{ingredsnr}.  We note that the binary parameter $A$
(column 11 and equation \ref{eq:Ia_rate}) is a function of the 
IMF chosen, ranging from $\sim 0.01$ for
the flat AY87
IMF to $\sim 0.15$ for the steep S86 IMF.  These
values ensure that the predicted present-day Type Ia SNe rate (column 12)
is $R_{\rm Ia}\approx 0.05\pm 0.03$ SNu (\cite{TCB94}).

\begin{table*}[hpt]
\caption[$M_{\rm g}(0)=10^{12}$ M$_\odot$ Model Sensitivity to IMF]
{\label{tbl:imf_tbl} Influence of IMF
selection upon the galactic wind time $t_{\rm
GW}$ (and hence, gas, O, and Fe, ejection, as well as resultant photometric
properties) for two different SNR thermal energy formalisms (Section 
\ref{ingredenergy}), for the $M_{\rm g}(0)=10^{12}$ M$_\odot$ model.  
``IMF'' refers to the mass function source:  S55=\cite{S55};
AY87=\cite{AY87}; KTG93=\cite{KTG93}; S86=\cite{Sc86}.  $A$ is the binary
parameter (equation \ref{eq:Ia_rate}) 
necessary to recover present-day Type Ia SNe rate.
All other ingredients as described in Section \ref{template}.
Superscript $\ast$ implies
that $\nu$ has been adjusted from that shown in Table \ref{tbl:templatemodels2}
in order to recover the appropriate
present-day photo-chemical properties 
(specifically, (V-K) and $[<{\rm Z}>]_{\rm V}$).
}
\begin{center}
\begin{tabular}{crccccrccrccr}
\vspace{2.0mm}
IMF & $\nu$$\;\;\;$ & $t_{\rm GW}$ & $m_{\rm g}^{\rm ej}$ &
$m_{\rm O}^{\rm ej}$ & $m_{\rm Fe}^{\rm ej}$ & M$_{\rm V}$ & B-V & V-K & 
$[<{\rm Z}>]_{\rm V}$ & $A$ & $R_{\rm Ia}$ & $<[{\rm Mg/Fe}]>_{\rm V}$\\
\multicolumn{13}{c}{\it SNe Thermal Energy Model B$_2$\rm} \\
S55$^\ast$ &  17.3 & 0.440 & 1.7e10 & 4.9e8 & 7.5e7 & -23.45 & 0.92 & 3.33 & +0.44$\;\;\;\;$ & 0.030 & 0.05 & +0.13$\;\;\;\;\;\;\;\;$ \\
AY87       &  17.3 & 0.563 & 3.5e10 & 2.3e9 & 3.4e8 & -23.08 & 0.95 & 3.97 & +0.77$\;\;\;\;$ & 0.012 & 0.05 & +0.31$\;\;\;\;\;\;\;\;$ \\
AY87$^\ast$&  88.3 & 0.049 & 9.6e10 & 6.6e9 & 6.2e8 & -23.03 & 0.83 & 3.38 & +0.44$\;\;\;\;$ & 0.012 & 0.03 & +0.34$\;\;\;\;\;\;\;\;$ \\
KTG93      &  17.3 & 0.450 & 1.2e10 & 1.7e8 & 4.9e7 & -23.90 & 0.86 & 2.96 & +0.20$\;\;\;\;$ & 0.060 & 0.06 & -0.10$\;\;\;\;\;\;\;\;$ \\
KTG93$^\ast$&  3.8 & 4.180 & 7.0e9  & 1.0e8 & 5.5e7 & -23.90 & 0.90 & 3.17 & +0.33$\;\;\;\;$ & 0.060 & 0.08 & -0.38$\;\;\;\;\;\;\;\;$ \\
S86        &  17.3 & 0.353 & 9.3e9  & 1.1e8 & 5.4e7 & -23.98 & 0.82 & 2.67 & -0.01$\;\;\;\;$ & 0.150 & 0.06 & -0.21$\;\;\;\;\;\;\;\;$ \\
\multicolumn{13}{c}{\it SNe Thermal Energy Model A$_0$\rm} \\
S55$^\ast$ &  17.3 & 0.453 & 1.6e10 & 4.6e8 & 7.2e7 & -23.45 & 0.92 & 3.34 & +0.44$\;\;\;\;$ & 0.030 & 0.05 & +0.13$\;\;\;\;\;\;\;\;$ \\
AY87       &  17.3 & 0.445 & 5.1e10 & 3.4e9 & 4.7e8 & -23.05 & 0.95 & 3.94 & +0.73$\;\;\;\;$ & 0.012 & 0.05 & +0.31$\;\;\;\;\;\;\;\;$ \\
AY87$^\ast$&  65.2 & 0.048 & 1.6e11 & 9.8e9 & 8.0e8 & -22.93 & 0.85 & 3.39 & +0.32$\;\;\;\;$ & 0.012 & 0.04 & +0.32$\;\;\;\;\;\;\;\;$ \\
KTG93      &  17.3 & 0.579 & 9.1e9  & 1.1e8 & 4.3e7 & -23.92 & 0.86 & 2.98 & +0.25$\;\;\;\;$ & 0.060 & 0.06 & -0.15$\;\;\;\;\;\;\;\;$ \\
KTG93$^\ast$&  7.7 & 2.418 & 5.5e9  & 7.5e7 & 4.8e7 & -23.95 & 0.89 & 3.13 & +0.31$\;\;\;\;$ & 0.060 & 0.07 & -0.37$\;\;\;\;\;\;\;\;$ \\
S86        &  17.3 & 0.467 & 5.7e9  & 5.1e7 & 4.3e7 & -23.99 & 0.82 & 2.69 & +0.10$\;\;\;\;$ & 0.150 & 0.06 & -0.31$\;\;\;\;\;\;\;\;$ \\
\end{tabular}
\end{center}
\end{table*}

A primary conclusion to be gleaned from inspection of Table \ref{tbl:imf_tbl}
is that the galactic wind time $t_{\rm GW}$ (column 3)
is not particularly sensitive to the
IMF, \it provided \rm
the star formation efficiency $\nu$ is kept constant (\ie
$\nu=17.3$ Gyr$^{-1}$), with $t_{\rm GW}$ ranging from $0.35\rightarrow 0.57$
Gyr.  

A secondary point of interest can be inferred from the S55 and AY87
entries to the Model B$_2$ section of Table \ref{tbl:imf_tbl}.  
One might naively expect that the flatter IMF (AY87) would \it always \rm
lead to an
earlier wind time because of its greater proportion of Type II SNe, whereas we
found that $t_{\rm GW}$ occurred $\sim 25$\% later for the flatter AY87 IMF.
This somewhat surprising behaviour can be traced to the metallicity dependence
of the SN shell cooling time and ISM merging time (Section \ref{ingredsnr})
in Cioffi \etal's (1988) evolutionary formalism.  Metallicity terms were not
considered by Cox (1972) and Chevalier (1974), which is why this behaviour has
not been encountered in previous models (nor in the Model A$_0$ S55 and AY87
entries to Table \ref{tbl:imf_tbl}), which have \it all \rm been based upon
these earlier supernova models.  A detailed analysis of this unexpected
behaviour is forthcoming, although a preliminary accounting can be found in
Gibson (1995).

While $t_{\rm GW}$ may not be overly IMF-sensitive, because of the different
proportion of low mass stars in the IMFs (\ie those which can effectively
lock-up and remove gas from possible subsequent enrichment), columns 4 through
6 show graphically that the predicted ejecta's mass and abundance can be.
For example, the S86 and AY87 IMFs, using SNR Model
A$_0$ and $\nu=17.3$ Gyr$^{-1}$, lead to almost identical $t_{\rm GW}$, but the
latter model predicts $\sim 9$ times the gas mass at $t_{\rm GW}$, and $\sim
70$ and $\sim 11$ times the mass of oxygen and iron, respectively.  This last
point is an interesting one -- recall that [O/Fe]$^{\rm ICM}$ lies in
the range $\sim +0.1\rightarrow +0.7$ (\cite{Mush94}) -- the models just
mentioned lead to ejecta with [O/Fe] of $\sim +0.05$ (the AY87 IMF)
and $\sim -0.74$ (the S86 IMF).  Obviously we must fold in a cluster
luminosity function before claiming anything, but the fact that one of the IMFs
leads to ejecta which is almost a full dex outside the observed ICM [O/Fe]
should be a clue to the anticipated difficulty in replicating the observations
using the steeper IMFs,
a point which we addressed in Gibson \& Matteucci (1997).

This iron overabundance relative to $\alpha$-elements in the ejecta of the S86
model is of course due to the
decreased significance of the Type II SNe in the steeper IMFs.  Type Ia SNe
play a correspondingly bigger role as the increased binary parameter $A$ would
suggest.  Not only does the wind
ejecta suffer from this Fe enhancement, but the
predicted luminosity-weighted [Mg/Fe] of the stellar population has a similar
problem.  The observed ratio in giant ellipticals is $\sim +0.2\rightarrow +0.3$
(\cite{WFG92}), in line with the AY87 IMF predictions of Table
\ref{tbl:imf_tbl}.  The S86 and KTG93 values are in the range
$\sim -0.2\rightarrow -0.4$, well outside the observations, another argument
against steep IMFs in ellipticals.

Studying the $\nu=17.3$ Gyr$^{-1}$ entries in Table \ref{tbl:imf_tbl}, we can
see that CML predictions for the S55 IMF
match the mean of the observations (as this is inherent in the template models
of Section \ref{template}).  The flatter IMF, because of its enhanced
enrichment history, by the same $t_{\rm GW}$, has redder and more metal-rich
stellar populations, at the present time.  Conversely, the steeper IMFs are
significantly bluer and metal poor relative to the mean of the observations.
For example, the KTG93 $\nu=17.3$ Gyr$^{-1}$ models are $\sim 0.4$ mag
too blue (in V-K) and $\sim 0.2$ dex metal-deficient, whereas the AY87
models are $\sim 0.6$ mag too red and $\sim 0.3$ dex too rich.  

To remedy this, we show a series of models with $\nu$ varied in order to best
replicate the present-day photo-chemical properties of the ellipticals.  These
models are represented with a $\ast$ in column 1 of Table \ref{tbl:imf_tbl}. 
For the flatter AY87 IMF, this means increasing $\nu$ by a factor of four
to five, leading to a much earlier wind ($t_{\rm GW}\approx 0.05$ Gyr as
opposed to $\sim 0.5$ Gyr).  With the earlier wind comes $\sim 3$ times the
mass of gas ejected at $t_{\rm GW}$, and a more extreme [O/Fe]$\approx
+0.2\rightarrow +0.3$.

An even more important result is found when we look at the attempt to match the
KTG93 and S86 IMF models with the observations.  Specifically, it
was found to be impossible to redden (or conversely, enrich) these
models to the observed mean of V-K=3.33 ($[<{\rm Z}>]_{\rm V}=+0.44$)
(for the luminosities involved here).
Even reducing $\nu$ by factors of four to five could only redden the colours by
$\sim 0.2$ mag, which is still $\sim 0.2$ mag blueward of the mean.  By this
point the wind epoch has shifted to $t\simgt 4$ Gyr, which for the assumed
cosmology would imply active observable star formation ($\psi\simgt 150$
M$_\odot$/yr) at redshifts $z\simlt 0.9$, contrary to observations 
(\cite{S86}).   By this
time the [Mg/Fe] of the stellar population has been decreased from an already
untenable $\sim -0.1$, to $\sim -0.4$, even further removed from the observed
overabundance of magnesium relative to iron.  We haven't shown the S86
predictions as they are even worse than the KTG93 ones.

This is an important result, and one which cannot be remedied within the
closed-box formalism for chemical evolution currently adopted in {\bf MEGaW}.
For these steep IMFs, the metallicity increases slowly because of the lack
of high mass stars in the IMF (at least, ``slower'' than for the S55
and AY87 IMFs).  Because of this lack of super metal rich stars
(compared to that seen with the flatter IMFs), all the stars formed 
with Z=0.0 (\ie all those formed prior to the first SNe explosions -- 
$t\simlt 0.03$ Gyr)
tend to ``pull'' the colours/metallicities too blue/low.\footnote{This is a
parallel manifestation to that of
the ``G-dwarf problem'' in the Milky Way, and
one from which all closed-box models of elliptical galaxy evolution suffer
(Worthey \etal 1996).}
This would seem to
indicate that this is partly an artifact of the star formation
formalism.  An infall model in which the mass of the gaseous component
increases with time according to some accretion timescale (as opposed to the
closed-box model in which all the gas is present and available for star
formation at $t=0$) may be an appropriate mechanism to ``save'' the steeper
IMFs (\eg \cite{TCBF95}), although the difficulty in
reconciling the [Mg/Fe] may still be problematic.\footnote{On the other hand,
independent supporting evidence \it against \rm steeper-than-Salpeter (1955) 
IMFs has already been put forth by Matteucci \& Gibson (1995), Zepf \& Silk
(1996) and Loewenstein \& Mushotzky (1996), based upon the significant
$\alpha$-element overabundances, with respect to iron, in the hot X-ray
emitting intracluster gas.}
Infall models will be
investigated in a future paper.

More complex scenarios involving bimodal epochs of star formation and IMFs are
a separate issue, and considered in Elbaz \etal (1995) and Gibson (1996a).

\subsubsection{Star formation rate}
\label{analsfr}

It is apparent from Figures \ref{fig:Zsfr} and \ref{fig:VKsfr} that
there is considerable scatter about the mean of the observed
CML relation.  A good part of this scatter is
intrinsic (\ie beyond observational error -- \cite{F77}).  One obvious source 
of scatter can most likely be traced to the inherently simplistic
handling of star formation in {\bf MEGaW}.  

To illustrate the effect of varying the star formation efficiency, we
ran models with efficiency parameters $\nu$ (equation \ref{eq:sfr}) arbitrarily
scaled up or down by a factor two, as compared with the template values for
$\nu$ listed in Table \ref{tbl:templatemodels2}.  
How this impacts upon the galactic wind
time, the mass and abundance of the ejected ISM, and the resultant present-day
photo-chemical properties of the remaining stellar population, can be seen in
Table \ref{tbl:tbl_sfr}, and graphically in Figures \ref{fig:Zsfr} and
\ref{fig:VKsfr}.

\begin{table*}[hpt]
\caption[Sensitivity to Star Formation Efficiency]{\label{tbl:tbl_sfr}
Sensitivity of $t_{\rm GW}$, and the resultant present-day
photo-chemical properties of ellipticals, to the star formation efficiency
parameter $\nu$.  The template model of Table \ref{tbl:templatemodels2} is 
given first.  Arbitrary scaling of $\nu$ up and down by factors of two are
listed subsequently.  All other input ingredients as discussed in Section
\ref{template}.
}
\begin{center}
\begin{tabular}{crccccrccr}
\vspace{2.0mm}
$M_{\rm g}(0)$ & $\nu$$\;\;\;$ & $t_{\rm GW}$ & $m_{\rm g}^{\rm ej}$ &
$m_{\rm O}^{\rm ej}$ & $m_{\rm Fe}^{\rm ej}$ & M$_{\rm V}$ & B-V & V-K & 
$[<{\rm Z}>]_{\rm V}$ \\
\multicolumn{10}{c}{\it Template $\nu$\rm} \\
1.0e6  & 188.9 & 0.006 & 3.5e5  & 2.2e2 & 1.0e1 &  -8.21 & 0.68 & 2.08 & -2.28$\;\;\;\;$ \\
5.0e7  & 209.7 & 0.007 & 1.2e7  & 3.2e4 & 1.5e3 & -12.61 & 0.69 & 2.12 & -1.56$\;\;\;\;$ \\
1.0e9  & 123.1 & 0.016 & 1.7e8  & 3.0e6 & 2.0e5 & -15.89 & 0.74 & 2.44 & -0.51$\;\;\;\;$ \\
5.0e10 &  46.0 & 0.077 & 3.4e9  & 1.1e8 & 9.3e6 & -20.15 & 0.85 & 3.04 & +0.13$\;\;\;\;$ \\
1.0e12 &  17.3 & 0.440 & 1.7e10 & 4.9e8 & 7.5e7 & -23.45 & 0.92 & 3.33 & +0.44$\;\;\;\;$ \\
\multicolumn{10}{c}{\it Template $\nu\times 2$\rm} \\
1.0e6  & 375.9 & 0.004 & 2.2e5  & 7.5e-2& 1.0e-5&  -8.38 & 0.68 & 2.07 & -3.23$\;\;\;\;$ \\
5.0e7  & 419.5 & 0.004 & 8.5e6  & 1.7e2 & 3.9e-4& -12.69 & 0.68 & 2.07 & -3.22$\;\;\;\;$ \\
1.0e9  & 246.3 & 0.008 & 1.7e8  & 8.4e5 & 4.7e4 & -15.94 & 0.70 & 2.15 & -1.25$\;\;\;\;$ \\
5.0e10 &  92.1 & 0.030 & 5.1e9  & 1.5e8 & 1.5e7 & -20.17 & 0.78 & 2.73 & -0.11$\;\;\;\;$ \\
1.0e12 &  34.5 & 0.164 & 2.7e10 & 8.4e8 & 9.9e7 & -23.44 & 0.87 & 3.19 & +0.34$\;\;\;\;$ \\
\multicolumn{10}{c}{\it Template $\nu\times 1/2$\rm} \\
1.0e6  &  94.0 & 0.010 & 4.1e5  & 1.3e3 & 5.4e1 &  -8.47 & 0.71 & 2.17 & -1.37$\;\;\;\;$ \\
5.0e7  & 104.9 & 0.014 & 1.3e7  & 1.4e5 & 7.6e3 & -12.54 & 0.74 & 2.34 & -0.76$\;\;\;\;$ \\
1.0e9  &  61.6 & 0.032 & 1.8e8  & 3.8e6 & 3.4e5 & -15.82 & 0.80 & 2.73 & -0.25$\;\;\;\;$ \\
5.0e10 &  23.0 & 0.202 & 2.2e9  & 6.8e7 & 6.9e6 & -20.99 & 0.90 & 3.23 & +0.29$\;\;\;\;$ \\
1.0e12 &   8.6 & 1.145 & 1.2e10 & 3.4e8 & 7.1e7 & -23.48 & 0.96 & 3.43 & +0.48$\;\;\;\;$ \\
\end{tabular}
\end{center}
\end{table*}

For all the models, the increased efficiency parameter (\ie $2\nu$ models)
leads to an increased SNe rate, which
for the more massive models (\ie $M_{\rm g}(0)\simgt 10^9$ M$_\odot$) results
in earlier galactic winds, with a correspondingly greater mass of gas and
metals ejected (up to $\sim 50$\% more).  For the lower mass (\ie dwarf)
models, the wind still occurs earlier, but there is less mass ejected because
there is a $\sim 4$ Myr delay before the most massive Type II SNe progenitors
explode -- the increased SFR during this ``delay'' phase means less gas is
available for expulsion at $t_{\rm GW}$.  The opposite behaviour is seen for
the decreased efficiency parameter (\ie $0.5\nu$ models).

Figures \ref{fig:Zsfr} and \ref{fig:VKsfr} show the scatter in the
observational planes for the factor of two variations in $\nu$.  The majority
of the observed data points fall within the $0.5\nu$ and $2\nu$ curves.
Obviously there are many uncertainties, but we do agree with 
Arimoto \& Yoshii (1987) that
much of the scatter in the observed photo-chemical correlations may be
attributable to intrinsic scatter in the star formation efficiency.

\begin{figure}[ht]
\centering
\vspace{8.5cm}
\includegraphics{`cat}
\caption[]{Solid curve represents the predicted
[Z]-M$_{\rm V}$ relation for the template
models of Table \ref{tbl:templatemodels2}.  The effects of doubling or halving
the star formation efficiency parameter $\nu$, from the template values, are
also shown. Observational data from Smith (1985) (open circles) and Terlevich
\etal (1981) (filled circles).
\label{fig:Zsfr}}
\end{figure}

\begin{figure}[ht]
\centering
\vspace{8.5cm}
\includegraphics{`cat}
\caption[]{Solid curve represents the predicted 
(V-K)-M$_{\rm V}$ relation for the template
models of Table \ref{tbl:templatemodels2}.  The effects of doubling or halving
the star formation efficiency parameter $\nu$, from the template values, are
also shown.  Observational data from Thuan (1985) (open boxes) and Bower \etal
(1992) (circles).
\label{fig:VKsfr}}
\end{figure}

\section{Summary}
\label{summary}

We have described, in some detail, the first version of {\bf MEGaW}, our
coupled photometric, chemical, and ISM thermal evolution code, based upon the
classic framework of Larson (1974b).  We have
attempted to step through, one by one, 
each of the primary ingredients, in order to demonstrate how the timeframe for
bulk star formation cessation (\ie $t_{\rm GW}$), as well as the resultant
predicted
present-day photo-chemical properties, changes, when adjusting any one of
these input parameters amongst what appear to be
several plausible \it a priori \rm selections.  The sensitivity of the results
to any one parameter has been hidden in previous studies of this sort.

We do not claim to have performed the strictest of statistical studies; the
analysis shown is meant to be illustrative, more than anything.  What we can
conclude at this point though is:

\begin{itemize}
\item{} The present star formation rate formalism ($\psi\propto M_{\rm
g}$) precludes the use of IMFs steeper-than-Salpeter (1955) (Section
\ref{analimf}).
\item{} Discriminating between early ($t_{\rm GW}\simlt 0.1$ Gyr) and
late ($t_{\rm GW}\approx 0.5$ Gyr) galactic winds via photo-chemical constraints
alone is not possible.  For a given assumption regarding the efficiency of SNR
energy transfer to the ISM, we can usually recover the present-day observations
by varying the star formation efficiency parameter appropriately (Section
\ref{analsnr}; \cite{GM96}).
\item{} For non-extreme distributions, the role played by dark matter
in setting $t_{\rm GW}$ would appear to be less important than at first
envisaged by Matteucci (1992) (Section \ref{analenergy}).
\item{} An inverse wind phenomenon has been observed whereby $t_{\rm
GW}$ actually increases with increasingly flatter IMFs.  This comes about
because of the metallicity dependence in the SNR evolution models of Cioffi
\etal (1988), which tends to reduce the effective energy transferred per SN
event to the ISM (Section \ref{analimf}).
\end{itemize}

An improved second version of {\bf MEGaW} is still under development.
Many enhancements to the basic code are
currently underway, including most importantly, a full hydrodynamical treatment
of the elliptical's evolution (\eg \cite{DFJ91}; \cite{CDPR91}).
A 1D-hydro code, based upon Thomas (1988), with the addition
of diffuse dark halos, is
being meshed with our photo-chemical evolution code.  Coupled with
more sophisticated radial star formation history scenarios, we will be
exploring the origin and evolution of photo-chemical gradients in ellipticals.
Other enhancements for Version 2 of {\bf
MEGaW} include a simple
multi-phase ISM treatment (\cite{FP93}), the inclusion of
a gas infall term (\cite{TCBF95}), and replacing the photometric evolution
module with a full spectral synthesis one.
A number of other minor enhancements were laid out in Gibson (1995).

Subsequent papers in this series will examine the evolution of
galaxy cluster ICM abundances, damped Lyman-$\alpha$ systems, and stellar
yield compilation differences.

\section*{ACKNOWLEDGEMENTS}
First, and foremost,
I would like to thank Francesca Matteucci for her boundless
patience and assistance.
Thanks also to Stan Woosley, Norbert Langer, and Friedel
Thielemann for providing nucleosynthesis data prior to publication.
Frank Timmes' help during the debugging process is gratefully acknowledged,
as is that of Guy Worthey's.
The consideration shown by the Astrophysics Department at Oxford during my
two-year ``sabbatical'' cannot be emphasised enough.  NSERC, Paul Hickson,
and the University of British Columbia,
are acknowledged for their respective financial assistance.

%%%%%%%%%%%%%%%%%%%%%%%%%%%%%%%%%%%%%%%%%%%%%%%%%%%%%%%%%%%%%%%%%%%%%%%%%%%%%%

%%%%%%%%%%%%%%%%%%%%%%%%%%%%%%%%%%%%%%%%%%%%%%%%%%%%%%%%%%%%%%%%%%%%%%%%%%%%%%%


\begin{thebibliography}{}

\bibitem[Alcock \& Paczynski 1978]{AP78}
Alcock, C. \& Paczynski, B. 1978,
\apj, 223, 244

\bibitem[Angeletti \& Giannone 1990]{AG90}
Angeletti, L. \& Giannone, P. 1990,
\aap, 234, 53 (AG90)

\bibitem[Angeletti \& Giannone 1991]{AG91}
Angeletti, L. \& Giannone, P. 1991,
\aap, 248, 45

\bibitem[Arimoto 1989]{A89}
Arimoto, N. 1989,
Evolutionary Phenomena in Galaxies, ed. J.E. Beckman \& B.E.J. Pagel,
Cambridge: Cambridge Univ. Press, 323

%\bibitem[Arimoto 1996]{Ar96}
%Arimoto, N. 1996,
%New Light on Galaxy Evolution, ed. R. Bender \etal, Dordrecht: Reidel, in press

%\bibitem[Arimoto \& Yoshii 1986]{AY86}
%Arimoto, N. \& Yoshii, Y. 1986,
%\aap, 164, 260

\bibitem[Arimoto \& Yoshii 1987]{AY87}
Arimoto, N. \& Yoshii, Y. 1987,
\aap, 173, 23  (AY87)

\bibitem[Arimoto \& Yoshii 1989]{AY89}
Arimoto, N. \& Yoshii, Y. 1989,
\aap, 224, 361)

%\bibitem[Arnaud 1994]{A94}
%Arnaud, M. 1994,
%Clusters of Galaxies, ed. F. Durret, A. Mazure \& J. Tran Thanh Van,
%Gif-sur-Yvette: Editions Frontieres, 211

\bibitem[Arnaud \etal 1992]{ARBVV92}
Arnaud, M., Rothenflug, R., Boulade, O., Vigroux, L. \& Vangioni-Flam, E. 1992,
\aap, 254, 49

\bibitem[Arnett 1978]{A78}
Arnett, W.D. 1978,
\apj, 219, 1008

\bibitem[Arnett 1991]{A91}
Arnett, D. 1991,
Frontiers of Stellar Evolution, ed. D.L. Lambert, ASP Conf. Series, 389  (A91)

%\bibitem[Arnett 1995a]{A95}
%Arnett, D. 1995a,
%private communication

%\bibitem[Arnett 1995b]{Arnett95}
%Arnett, D. 1995b,
%\araa, 33, 115

\bibitem[Arnett 1996]{A96}
Arnett, D. 1996,
Supernovae and Nucleosynthesis, Princeton: Princeton Univ. Press

%\bibitem[Arnett 1996b]{Dave96}
%Arnett, D., 1996b,
%Formation of the Galactic Halo, ed. H. Morrison \& A. Sarajedini, ASP Conf.
%Series, 337

\bibitem[Babul \& Rees 1992]{BR92}
Babul, A. \& Rees, M.J. 1992,
\mnras, 255, 346

%\bibitem[Bachcall \etal 1994]{BFGK94}
%Bahcall, J.N., Flynn, C., Gould, A. \& Kirhakos,S. 1994,
%\apj, 435, L51

%\bibitem[Barbaro \& Olivi 1989]{BO89}
%Barbaro, G. \& Olivi, F.M. 1989,
%\apj, 337, 125

%\bibitem[Barbuy 1994]{B94}
%Barbuy, B. 1994,
%\apj, 430, 218

\bibitem[Baum 1959]{B59}
Baum, W.A. 1959,
\pasp, 71, 106

\bibitem[Bazan \& Mathews 1990]{BM90}
Bazan, G. \& Mathews, G.J. 1990,
\apj, 354, 644

\bibitem[Becker 1981]{B81}
Becker, S.A. 1981,
\apjs, 45, 475

%\bibitem[Bender \etal 1992]{BBF92}
%Bender, R., Burstein, D. \& Faber, S.M. 1992,
%\apj, 399, 462

%\bibitem[Bernstein \etal 1995]{BWNUT95}
%Bernstein, G.M., Nichol, R.C., Tyson, A.A., Ulmer, M.P. \& Wittman, D. 1995,
%\aj, 110, 1507

\bibitem[Bertin \etal 1992]{BSS92}
Bertin, G., Saglia, R.P. \& Stiavelli, M. 1992,
\apj, 384, 423

\bibitem[Bertin \etal 1994]{BBB94}
Bertin, G., Bertola, F., Buson, L.M., Danziger, I.J., Dejonghe, H., Sadler,
E.M., Saglia, R.P., de Zeeuw, P.T. \& Zeilinger, W.W. 1994, 
\aap, 292, 381

\bibitem[Bertelli \etal 1994]{BBCFN94}
Bertelli, G., Bressan, A., Chiosi, C., Fagotto, F. \& Nasi, E. 1994,
A\&AS, 106, 275

%\bibitem[Bessell \etal 1991]{BBSW91}
%Bessell, M.S., Brett, J.M., Scholz, M. \& Wood, P.R. 1991,
%A\&AS, 89, 335

%\bibitem[Binggeli \etal 1987]{BST87}
%Binggeli, B., Sandage, A. \& Tammann, G.A. 1987, 
%\aj, 94, 251

%\bibitem[Binney \& Tremaine 1987]{BT87}
%Binney, J. \& Tremaine, S. 1987,
%Galactic Dynamics, Princeton Univ. Press, Princeton

\bibitem[Bland-Hawthorn 1995]{Bland95}
Bland-Hawthorn, J. 1995,
\pasa, 12, 190

%\bibitem[Boothroyd 1994]{Booth94}
%Boothroyd, A. 1994,
%private communication

\bibitem[Bower \etal 1992]{BLE92}
Bower, R.G., Lucey, J.R. \& Ellis, R.S. 1992,
\mnras, 254, 589

\bibitem[Bressan \etal 1994]{BCF94}
Bressan, A., Chiosi, C. \& Fagotto, F. 1994,
\apjs, 94, 63

%\bibitem[Brocato \etal 1990]{BMMT90}
%Brocato, E., Matteucci, F., Mazzitelli, I. \& Tornamb\`e, A. 1990,
%\apj, 349, 458

%\bibitem[Bruzual 1983]{B83}
%Bruzual A.,G. 1983,
%\apj, 273, 105

\bibitem[Bruzual \& Charlot 1993]{BC93}
Bruzual A., G. \& Charlot, S. 1993,
\apj, 405, 538

%\bibitem[Burkert \& Hensler 1989]{BH89}
%Burkert, A. \& Hensler, G. 1989,
%Evolutionary Phenomena in Galaxies, ed. J.E. Beckman \& B.E.J. Pagel,
%Cambridge: Cambridge Univ. Press, 230

%\bibitem[Buzzoni 1989]{B89}
%Buzzoni, A. 1989,
%\apjs, 71, 817

%\bibitem[Buzzoni 1995]{B95}
%Buzzoni, A. 1995,
%\apjs, 98, 101

%\bibitem[Canizares \etal 1988]{CMD88}
%Canizares, C.R., Markert, T.H. \& Donahue, M.E. 1988, 
%Cooling Flows in Clusters and Galaxies, ed. A.C. Fabian,
%Dordrecht: Kluwer, 63

%\bibitem[Cannon \etal 1995]{CFLW95}
%Cannon, R.C., Frost, C.A., Lattanzio, J.C. \& Wood, P.R. 1995,
%preprint

\bibitem[Carigi 1994]{C94}
Carigi, L. 1994,
\apj, 424, 181

%\bibitem[Carigi \etal 1995]{CCPS95}
%Carigi, L., Col\'in, P., Peimbert, M. \& Sarmiento, A. 1995,
%\apj, 445, 98

%\bibitem[Carollo \& Danziger 1993]{CD93}
%Carollo, C.M. \& Danziger, I.J. 1993, 
%\mnras, 265, 553

%\bibitem[Caughlan \& Fowler 1988]{CF88}
%Caughlan, G.R. \& Fowler, W.A. 1988,
%At. Data Nucl. Data Tables, 40, 283

%\bibitem[Caughlan \etal 1985]{CFHZ85}
%Caughlan, G.R., Fowler, W.A., Harris, M.J. \& Zimmerman, B.A. 1985,
%At. Data Nucl. Data Tables, 32, 197

%\bibitem[Charlot \& Bruzual 1991]{CB91}
%Charlot, A. \& Bruzual A., G. 1991,
%\apj, 367, 126

%\bibitem[Charlot \etal 1996]{CWB95}
%Charlot, S., Worthey, G. \& Bressan, A. 1996,
%\apj, 457, 625

\bibitem[Chevalier 1974]{C74}
Chevalier, R.A. 1974,
\apj, 188, 501

%\bibitem[Chuvenkov \& Glukhov 1995]{CG95}
%Chuvenkov, V. \& Glukhov, A. 1995,
%The Light Element Abundances, ed. P. Crane, Berlin: Springer-Verlag, 124

\bibitem[Cioffi \etal 1988]{CMB88}
Cioffi, D.F., McKee, C.F. \& Bertschinger, E. 1988,
\apj, 334, 252

\bibitem[Ciotti \etal 1991]{CDPR91}
Ciotti, L., d'Ercole, A., Pellegrini, S. \& Renzini, A. 1991,
\apj, 376, 380

%\bibiten[Cole \etal 1995]{CAFNZ95}
%Cole, S., Aragon-Salamanca, A., Frenk, C.S., Navarro, J.F. \& Zepf, S.E. 1995,
%\mnras, in press

%\bibitem[Couture \& Hardy 1990]{CH90}
%Couture, J. \& Hardy, E. 1990,
%\aj, 99, 540

%\bibitem[Couture \& Hardy 1993]{CH93}
%Couture, J. \& Hardy, E. 1993,
%\apj, 406, 142

\bibitem[Cox 1972]{C72}
Cox, D.P. 1972,
\apj, 178, 159

%\bibitem[David \etal 1990]{DFJ90}
%David, L.P., Forman, W. \& Jones, C. 1990,
%\apj, 359, 29

\bibitem[David \etal 1991]{DFJ91}
David, L.P., Forman, W. \& Jones, C. 1991,
\apj, 380, 39

%\bibitem[Davidge 1994]{D94}
%Davidge, T.J. 1994,
%\aj, 108, 2123

%\bibitem[Davidge \& Jones 1992]{DJ92}
%Davidge, T.J. \& Jones, J.H. 1992,
%\aj, 104, 1365

\bibitem[Dekel \& Silk 1986]{DS86}
Dekel, A. \& Silk, J. 1986,
\apj, 303, 39

%\bibitem[De Propris \etal 1995]{DPHM95}
%De Propris, R., Pritchet, C.J., Harris, W.E. \& McClure, R.D. 1995,
%\apj, 450, 534

%\bibitem[De Young \& Heckman 1994]{DH94}
%De Young, D.S. \& Heckman, T.M. 1994,
%\apj, 431, 598

%\bibitem[D\'iaz \& Tosi 1984]{DT84}
%D\'iaz, A.I. \& Tosi, M. 1984,
%\mnras, 208, 365

%\bibitem[Driver \etal 1994]{DPDMD94}
%Driver, S.P., Phillips, S., Davies, J.I., Morgan, I. \& Disney, M.J. 1994,
%\nat, 268, 393

\bibitem[Elbaz \etal 1995]{EAV95}
Elbaz, D., Arnaud, M. \& Vangioni-Flam, E. 1995,
\aap, 303, 345

\bibitem[Einsel \etal 1995]{EFKF95}
Einsel, Ch., Fritze-v. Alvensleben, U., Kr\"uger, H. \& Fricke, K.J. 1995,
\aap, 296, 347

%\bibitem[Faber \& Gallagher 1976]{FG76}
%Faber, S.M. \& Gallagher, J.S. 1976,
%\apj, 204, 365

\bibitem[Faber 1977]{F77}
Faber, S.M. 1977,
The Evolution of Galaxies and Stellar Populations, 
ed. B.M. Tinsley \& R.B. Larson,
New haven: Yale Obs., 157

\bibitem[Fagotto \etal 1994]{FBBC94}
Fagotto, F., Bressan, A., Bertelli, G. \& Chiosi, C. 1994,
\aaps, 104, 365

%\bibitem[Fan \& Tytler 1994]{FanT94}
%Fan, X.M. \& Tytler, D. 1994,
%\apjs, 94, 17

%\bibitem[Ferguson \& Binggeli 1994]{FB94}
%Ferguson, H.C. \& Binggeli, B. 1994,
%\aapr, 6, 67

%\bibitem[Ferguson \& Sandage 1995]{FS91}
%Ferguson, H.C. \& Sandage, A. 1991,
%\aj, 101, 765

\bibitem[Ferrini \& Poggianti 1993]{FP93}
Ferrini, F. \& Poggianti, B.M. 1993,
\apj, 410, 44

%\bibitem[Ferrini \etal 1992]{FMPP92}
%Ferrini, F., Matteucci, F., Pardi, C. \& Penco, U. 1992, 
%\apj, 387, 138

%\bibitem[Filippenko \& Sargent 1986]{FS86}
%Filippenko, A.V. \& Sargent, W.L.W. 1986,
%\aj, 91, 691

%\bibitem[Forestini \& Charbonnel 1996]{FC96}
%Forestini, M. \& Charbonnel, C. 1996,
%\aaps, in press

%\bibitem[Forman \etal 1985]{FJT85}
%Forman, W., Jones, C. \& Tucker, W. 1985,
%\apj, 293, 102

%\bibitem[Forman \etal 1994]{FJT94}
%Forman, W., Jones, C. \& Tucker, W. 1994,
%\apj, 429, 77

%\bibitem[Forman \etal 1993]{FJDFMO93}
%Forman, W., Jones, C., David, L., Fraux, L., Makishima, M. \& Ohashi, K. 1993, 
%\apj, 418, L75

%\bibitem[Franco \etal 1994]{FMATT94}
%Franco, J., Miller III, W.W., Arthur, S.J., Tenorio-Tagle, G., Terlevich, R.J.,
%1994,
%\apj, 435, 805

%\bibitem[Fran\c cois \& Matteucci 1993]{FM93}
%Fran\c cois, P. \& Matteucci, F. 1993, 
%\aap, 280, 136

%\bibitem[Fria\c ca \& Terlevich 1994]{FT94}
%Fria\c ca, A.C.S. \& Terlevich, R.J. 1994,
%preprint

\bibitem[Giovagnoli \& Tosi 1995]{GT95}
Giovagnoli, A. \& Tosi, M. 1995,
\mnras, 273, 499

\bibitem[Gibson 1994a]{G94a}
Gibson, B.K. 1994a,
\mnras, 271, L35

\bibitem[Gibson 1994b]{G94b}
Gibson, B.K. 1994b,
\jrasc, 88, 383

\bibitem[Gibson 1995]{G95a}
Gibson, B.K. 1995,
Ph.D. Thesis, University of British Columbia

\bibitem[Gibson 1996a]{G95b}
Gibson, B.K. 1996a,
\mnras, 278, 829

\bibitem[Gibson 1996b]{G96}
Gibson, B.K. 1996b,
\apj, 468, 167

\bibitem[Gibson \& Matteucci 1997]{GM96}
Gibson, B.K. \& Matteucci, F. 1997,
\apj, 475, 47

\bibitem[Gibson \& Mould 1997]{GM97}
Gibson, B.K. \& Mould, J.R. 1997,
\apj, 482, in press

%\bibitem[Gonz\'alez \& Gorgas 1995]{GG95}
%Gonz\'alez, J.J. \& Gorgas, J. 1995,
%Fresh Views on Elliptical Galaxies, ed. A. Buzzoni \& A.
%Renzini, ASP Conf. Series, p. 225

\bibitem[Greggio \& Renzini 1983]{GR83}
Greggio, L. \& Renzini, A. 1983,
\aap, 118, 217

\bibitem[Guiderdoni \& Rocca-Volmerange 1987]{GRV87}
Guiderdoni, B. \& Rocca-Volmerange, B. 1987,
\aap, 186, 1

\bibitem[G\"usten \& Mezger 1982]{GM82}
G\"usten, R. \& Mezger, P.G. 1982,
Vistas Astron., 26, 159

%\bibitem[Hamann \& Ferland 1993]{HF93}
%Hamann, F. \& Ferland, G. 1993,
%\apj, 418, 11

%\bibitem[Hardy \etal 1994]{HCCJ94}
%Hardy, E., Couture, J., Couture, C., Joncas, G. 1994,
%\aj, 107, 195

\bibitem[Hills 1980]{H80}
Hills, J.G. 1980,
\apj, 225, 986

%\bibitem[Iben 1991]{I91}
%Iben, I. 1991,
%\apjs, 76, 55

%\bibitem[Iben \& Tutukov 1987]{IT87}
%Iben, I. \& Tutukov, A.V. 1987,
%\apj, 311, 753

\bibitem[Ikeuchi 1977]{I77}
Ikeuchi, S. 1977,
Prog. Theor. Phys., 58, 1742

%\bibitem[Ishimaru \etal 1994]{ITAN94}
%Ishimaru, Y., Tsujimoto, T., Arimoto, N. \& Nomoto, K. 1994,
%Evolution of the Universe and its Observational Quest, ed. K.
%Sato, Tokyo: Universal Acad. Press, 457

\bibitem[Iwamoto \etal 1994]{INH94}
Iwamoto, K., Nomoto, K. \& Hashimoto, M. 1994,
Evolution of the Universe and its Observational
Quest, ed. K. Sato, Tokyo: Universal Academy Press, 459

%\bibitem[Johnson \& Axford 1971]{JA71}
%Johnson, H.E. \& Axford, W.I. 1971,
%\apj, 165, 381

%\bibitem[Kaastra \& Mewe 1993]{KM93}
%Kaastra, J.S. \& Mewe, R. 1993,
%A\&AS, 97, 443

%\bibitem[Kennicutt 1989]{K89}
%Kennicutt, R.C. 1989,
%\apj, 344, 685

%\bibitem[Kormendy 1990]{K90}
%Kormendy, J. 1990,
%Evolution of the Universe of Galaxies, ed. R.G. Kron, ASP Conf. Series, 33

\bibitem[Kroupa \etal 1993]{KTG93}
Kroupa, P., Tout, C.A. \& Gilmore, G. 1993,
\mnras, 262, 545  (KTG93)

\bibitem[Kurucz 1993]{K93}
Kurucz, R.L. 1993,
Kurucz CD-ROM No. 13, SAO, Cambridge

%\bibitem[Lacey \& Fall 1985]{LF85}
%Lacey, C.G. \& Fall, S.M. 1985,
%\apj, 290, 154

%\bibitem[Langer 1992]{L92}
%Langer, N. 1992,
%\aap, 265, L17

%\bibitem[Langer 1994]{L94}
%Langer, N. 1994,
%private communication

\bibitem[Langer \& Henkel 1995]{LH95}
Langer, N. \& Henkel, C. 1995,
\ssr, 74, 343  (LH95)

\bibitem[Larson 1974a]{L74a}
Larson, R.B. 1974a,
\mnras, 166, 585

\bibitem[Larson 1974b]{L74b}
Larson, R.B. 1974b,
\mnras, 169, 229

%\bibitem[Larson \& Dinerstein 1975]{LD75}
%Larson, R.B, Dinerstein, H.L. 1975,
%\pasp, 87, 911

%\bibitem[Leitherer \etal 1992]{LRD92}
%Leitherer, C., Robert, C. \& Drissen, L. 1992,
%\apj, 401, 596

%\bibitem[Liedahl \etal 1995]{LOG95}
%Liedahl, D.A., Osterheld, A.L. \& Goldstein, W.H. 1995,
%\apj, 438, L115

%\bibitem[Lipman \etal 1995]{LPH95}
%Lipman, K., Pettini, M. \& Hunstead, R.W. 1995,
%QSO Absorption Lines, ed. G. Meylan, Berlin: Springer-Verlag, in press

\bibitem[Loewenstein \& Mushotzky 1996]{LM96}
Loewenstein, M. \& Mushotzky, R.F. 1996,
\apj, 466, 695

%\bibitem[Maeder 1990]{M90}
%Maeder, A. 1990,
%A\&AS, 84, 139

\bibitem[Maeder 1992]{Md92}
Maeder, A. 1992,
\aap, 264, 105  (Erratum: 1993, \aap, 268, 833)  (M92)

%\bibitem[Maeder \& Conti 1994]{MC94}
%Maeder, A. \& Conti, P.S. 1994,
%\araa, 32, 227

\bibitem[Marigo \etal 1996]{MBC96}
Marigo, P., Bressan, A. \& Chiosi, C., 1996,
\aap, 313, 545

%\bibitem[Masai 1994]{Masai94}
%Masai, K. 1994,
%JQS\&RT, 51, 211

%\bibitem[Mathews 1989]{M89}
%Mathews, W.G. 1989,
%\aj, 97, 42

\bibitem[Mathews \& Baker 1971]{MB71}
Mathews, W.G. \& Baker, J.C. 1971,
\apj, 170, 241

\bibitem[Matteucci 1991]{M91}
Matteucci F. 1991,
Frontiers of Stellar Evolution, ed. D.L. Lambert, ASP Conf. Series, 539

\bibitem[Matteucci 1992]{M92}
Matteucci F. 1992,
\apj, 397, 32

\bibitem[Matteucci 1994]{M94}
Matteucci F. 1994,
\aap, 288, 57

\bibitem[Matteucci 1997]{M97}
Matteucci F. 1997,
Fund. Cosm. Phys., in press

%\bibitem[Matteucci 1995]{Matt95}
%Matteucci F. 1995,
%private communication

%\bibitem[Matteucci \& Fran\c cois 1989]{MF89}
%Matteucci, F. \& Fran\c cois, P. 1989, 
%\mnras, 239, 885

%\bibitem[Matteucci \& Gibson 1995]{MG95}
%Matteucci, F. \& Gibson, B.K. 1995,
%\aap, 304, 11

\bibitem[Matteucci \& Greggio 1986]{MG86}
Matteucci, F. \& Greggio, L. 1986,
\aap, 154, 279

%\bibitem[Matteucci \& Padovani 1993]{MP93}
%Matteucci, F. \& Padovani, P. 1993, 
%\apj, 419, 485

\bibitem[Matteucci \& Tornamb\`e 1987]{MT87}
Matteucci, F. \& Tornamb\`e, A. 1987,
\aap, 185, 51 (MT87)

%\bibitem[Matteucci \& Vettolani 1988]{MV88}
%Matteucci, F. \& Vettolani, G. 1988,
%\aap, 202, 21

%\bibitem[Matteucci \etal 1996]{MMV96}
%Matteucci, F., Molaro, P. \& Vladilo, G. 1996,
%\aap, in press

%\bibitem[Mazzei \etal 1994]{MdZX94}
%Mazzei, P., de Zotti, G. \& Xu, C. 1994,
%\apj, 422, 81

%\bibitem[McWilliam \& Rich 1994]{MR94}
%McWilliam, A. \& Rich, R.M. 1994,
%\apjs, 91, 749

%\bibitem[Melnick \etal 1977]{MWH77}
%Melnick, J., White, S.D.M. \& Hoessel, J. 1977,
%\mnras, 180, 207

%\bibitem[Meynet \etal 1994]{MMSSC94}
%Meynet, G., Maeder, A., Schaller, G., Schaerer, D. \& Charbonnel, C. 1994,
%A\&AS, 103, 97

%\bibitem[Michaud 1995]{M95}
%Michaud, G. 1995,
%private communication

\bibitem[Mihara \& Takahara 1994]{MT94}
Mihara, K. \& Takahara, F. 1994,
\pasj, 46, 447

%\bibitem[Minniti \etal 1995]{MOLWHI95}
%Minniti, D., Olszewski, E.W., Liebert, J., White, S.D.M., Hill, J.M. \& Irwin,
%M.J. 1995,
%\mnras, 277, 1293

%\bibitem[Missoulis 1994]{Miss94}
%Missoulis, V. 1994,
%Astr. Rep., 38, 12

%\bibitem[Mitchell \etal 1976]{MCDI76}
%Mitchell, R.J., Culhane, J.L., Davison, P.J. \& Ives, J.C. 1976,
%\mnras, 176, 29p

\bibitem[Mushotzky 1994]{Mush94}
Mushotzky, R. 1994,
ed. F. Durret, A. Mazure \& J. Tran Thanh Van,
Clusters of Galaxies, Gif-sur-Yvette: Editions Frontieres, 167

%\bibitem[Mushotzky \etal 1981]{MHSBH81}
%Mushotzky, R.F., Holt, S.S., Smith, B.W., Boldt, E.A. \& Holt, S.S. 1981, 
%\apj, 225, 21

\bibitem[Nath \& Chiba 1995]{NC95}
Nath, B.B. \& Chiba, M. 1995,
\apj, 454, 604

\bibitem[Nomoto \etal 1984]{NMY84}
Nomoto, K., Thielemann, F.K. \& Yokoi, K. 1984, 
\apj, 286, 644

%\bibitem[Oemler 1974]{O74}
%Oemler, A. 1974,
%\apj, 194, 1

\bibitem[Okazaki \etal 1993]{OCKF93}
Okazaki, Y., Chiba, M., Kumai, Y. \& Fujimoto, M. 1993,
\pasj, 45, 669

%\bibitem[Olofsson 1989]{O89}
%Olofsson, K. 1989,
%A\&AS, 80, 317

%\bibitem[Oort 1951]{O51}
%Oort, J.H. 1951,
%Problems of Cosmical Aerodynamics, Dayton: Central Air Documents Office, 118

%\bibitem[Ostriker \& McKee 1988]{OM88}
%Ostriker, J.P. \& McKee, C.F. 1988,
%Rev. Mod. Phys., 60, 1

%\bibitem[Padovani \& Matteucci 1993]{PM93}
%Padovani, P. \& Matteucci, F. 1993,
%\apj, 416, 26

%\bibitem[Paresce \etal 1995]{PDR95}
%Paresce, F., De Marchi, G. \& Romaniello, M. 1995,
%\apj, 440, 216

%\bibitem[Pastor \etal 1989]{PCSP89}
%Pastor, J., Canal, R., Sanahuja, B. \& Pell\'o, R. 1989,
%\apss, 157, 221

%\bibitem[Pilyugin 1993]{P93}
%Pilyugin, L.S. 1993,
%\aap, 277, 42

\bibitem[Prantzos \etal 1993]{PCV93}
Prantzos, N., Cass\'e, M. \& Vangioni-Flam, E. 1993,
\apj, 403, 630

%\bibitem[Rana \& Wilkinson 1988]{RW88}
%Rana, N.C. \& Wilkinson, D.A. 1988,
%\mnras, 231, 509

%\bibitem[Raymond \& Smith 1977]{RS77}
%Raymond, J.C. \& Smith, B.W. 1977,
%\apjs, 35, 419

\bibitem[Reimers 1975]{R75}
Reimers, D. 1975,
M\'em.~Soc.~Roy.~Sci.~Li\`ege, 8, 369

%\bibitem[Reimers \etal 1992]{R92}
%Reimers, D., Vogel, S., Hagen, H.-J., Engels, D., Groote, D., Wamsteker, W.,
%Clavel, J. \& Rosa, M.R. 1992,
%\nat, 360, 561

%\bibitem[Renzini \& Buzzoni 1986]{RB86}
%Renzini, A. \& Buzzoni, A. 1986,
%Spectral Evolution of Galaxies,
%ed. C. Chiosi \& A. Renzini, 
%Dordrecht: Reidel, 195

\bibitem[Renzini \& Voli 1981]{RV81}
Renzini, A. \& Voli, M. 1981,
\aap, 94, 175  (RV81)

\bibitem[Renzini \etal 1993]{RCDP93}
Renzini, A., Ciotti, L., d'Ercole, A. \& Pellegrini, S. 1993,
\apj, 419, 52

%\bibitem[Rich 1996]{R96}
%Rich, R.M. 1996,
%New Light on Galaxy Evolution, ed. R. Bender \etal, Dordrecht: Reidel, in press

%\bibitem[Rose 1985]{R85}
%Rose, J.A. 1985,
%\aj, 90, 1927

\bibitem[Saito 1979a]{S79a}
Saito, M. 1979a,
\pasj, 31, 181

\bibitem[Saito 1979b]{S79b}
Saito, M. 1979b,
\pasj, 31, 193

\bibitem[Saglia \etal 1992]{SBS92}
Saglia, R.P., Bertin, G. \& Stiavelli, M. 1992,
\apj, 384, 433

\bibitem[Salpeter 1955]{S55}
Salpeter, E.E. 1955,
\apj, 121, 161  (S55)

\bibitem[Sandage 1986]{S86}
Sandage, A. 1986,
\aap, 161, 89

%\bibitem[Sarazin 1979]{Sar79}
%Sarazin, C.L. 1979,
%\aplett, 20, 93

%\bibitem[Sarazin 1986]{Sar86}
%Sarazin, C.L. 1986,
%Rev. Mod. Phys., 58, 1

%\bibitem[Savaglio \& Webb 1995]{SW95}
%Savaglio, S. \& Webb, J. 1995,
%preprint

\bibitem[Scalo 1986]{Sc86}
Scalo, J.M. 1986,
Fund. Cosm. Phys., 11, 1  (S86)

\bibitem[Schaller \etal 1992]{SSMM92}
Schaller, G., Schaerer, D., Meynet, G. \& Maeder, A. 1992,
\aaps, 96, 269

%\bibitem[Schechter 1976]{S76}
%Schechter, P. 1976,
%\apj, 203, 297

\bibitem[Schmidt 1959]{S59}
Schmidt, M. 1959,
\apj, 129, 243

%\bibitem[Serlemitsos \etal 1993]{SLMMP93}
%Serlemitsos, P.J., Lowenstein, M., Mushotzky, R.F., Marshall, F.E., \& 
%Petre, R. 1993,
%\apj, 413, 518

%\bibitem[Shull \& Saken 1995]{SS95}
%Shull, J.M. \& Saken, J.M. 1995,
%\apj, 444, 663

%\bibitem[Sil'chenko 1994]{S94}
%Sil'chenko, O.K. 1994,
%\azh, 71, 7

%\bibitem[Smecker-Hane 1992]{S92}
%Smecker-Hane, T.A. 1992,
%Ph.D. Thesis, Johns Hopkins University

\bibitem[Smith 1985]{S85}
Smith, G.H. 1985,
\pasp, 97, 1058

%\bibitem[Spiegel 1981]{S81}
%Spiegel, M.R. 1981,
%Applied Differential Equations, 3rd Ed., Prentice-Hall, New Jersey

%\bibitem[Sutherland \& Dopita 1993]{SD93}
%Sutherland, R.S. \& Dopita, M.A. 1993,
%\apjs, 88, 253

\bibitem[Talbot \& Arnett 1971]{TA71}
Talbot, R.J. \& Arnett, W.D. 1971,
\apj, 170, 409

%\bibitem[Talbot \& Arnett 1973]{TA73}
%Talbot, R.J. \& Arnett, D.W. 1973,
%\apj, 186, 51

\bibitem[Tantalo \etal 1995]{TCBF95}
Tantalo, R., Chiosi, C., Bressan, A. \& Fagotto, F. 1995,
\aap, 311, 361

%\bibitem[Terlevich \& Melnick 1985]{TM85}
%Terlevich, R. \& Melnick, J. 1985,
%\mnras, 213, 841

\bibitem[Terlevich \etal 1981]{TDFB81}
Terlevich, R., Davies, R.J., Faber, S.M. \& Burstein, D. 1981,
\mnras, 196, 381

\bibitem[Thielemann \etal 1993]{TNH93}
Thielemann, F.-K., Nomoto, K. \& Hashimoto, M. 1993,
Origin and Evolution of the Elements, 
ed. N. Prantzos, E. Vangioni-Flam, \& M. Cass\'e,
Cambridge: Cambridge Univ. Press, 297

\bibitem[Thielemann \etal 1996]{TNH95}
Thielemann, F.-K., Nomoto, K. \& Hashimoto, M. 1996,
\apj, 460, 408 (TNH95)

%\bibitem[Thielemann \etal 1986]{TNY86}
%Thielemann, F.-K., Nomoto, K. \& Yokoi, K. 1986,
%\aap, 158, 17

\bibitem[Thomas 1988]{T88}
Thomas, P. 1988,
\mnras, 235, 315

\bibitem[Thuan 1985]{T85}
Thuan, T.X. 1985,
\apj, 299, 881

%\bibitem[Thuan \& Kormendy 1977]{TK77}
%Thuan, T.X. \& Kormendy, J. 1977,
%\pasp, 89, 466

%\bibitem[Timmes 1995]{T95}
%Timmes, F.X. 1995,
%private communication

\bibitem[Timmes \etal 1995]{TWW95}
Timmes, F.X., Woosley, S.E. \& Weaver, T.A. 1995,
\apjs, 98, 617

\bibitem[Tinsley 1980]{T80}
Tinsley, B.M. 1980,
Fund. Cosm. Phys., 5, 287

%\bibitem[Tinsley \& Larson 1979]{TL79}
%Tinsley, B.M. \& Larson, R.B. 1979, 
%\mnras, 186, 503

\bibitem[Tomisaka 1992]{T92}
Tomisaka, K. 1992,
\pasj, 44, 177

%\bibitem[Tornamb\'e \& Matteucci 1987]{TM87}
%Tornamb\'e, A. \& Matteucci, F. 1987,
%\apj, 318, L25

%\bibitem[Tornamb\'e 1989]{T89}
%Tornamb\'e, A. 1989,
%\mnras, 239, 771

%\bibitem[Tosi 1988]{T88}
%Tosi, M. 1988,
%\aap, 197, 33

\bibitem[Turatto \etal 1994]{TCB94}
Turatto, M., Cappellaro, E. \& Benetti, S. 1994,
\aj, 108, 202

\bibitem[van den Hoek \& Groenewegen 1997]{vG96}
van den Hoek, L.B. \& Groenewegen, M.A.T. 1997,
\aap, in press

\bibitem[Vader 1987]{V87}
Vader, J.P. 1987,
\apj, 317, 128

%\bibitem[VandenBerg \etal 1983]{VHDA83}
%VandenBerg, D.A, Hartwick, F.D.A., Dawson, P. \& Alexander, D.R. 1983,
%\apj, 266, 747

%\bibitem[Vladilo \etal 1995]{VDMS95}
%Vladilo, G., D'Odorico, S., Molaro, P. \& Savaglio, S. 1995,
%QSO Absorption Lines, ed. G. Meylan, Berlin: Springer-Verlag, 103

%\bibitem[Wang 1995]{Wang95}
%Wang, B.Q. 1995,
%\apj, 444, 590

%\bibitem[Weaver \etal 1977]{WMCSM77}
%Weaver, R., McCray, R., Castor, J., Shapiro, P. \& Moore, R. 1977,
%\apj, 218, 377

\bibitem[Weaver \& Woosley 1993]{WW93}
Weaver, T.A. \& Woosley, S.E. 1993,
\physrep, 227, 65

%\bibitem[Weiss \etal 1995]{WPM95}
%Weiss, A., Peletier, R.F. \& Matteucci, F. 1995,
%\aap, 296, 73

\bibitem[Whelan \& Iben 1973]{WI73}
Whelan, J.C. \& Iben Jr., I. 1973,
\apj, 186, 1007

%\bibitem[White 1991]{W91}
%White III, R.E. 1991,
%\apj, 367, 69

%\bibitem[White \etal 1994]{WDHH94}
%White III, R.E., Day, C.S.R., Hatsukade, I. \& Hughes, J.P. 1994,
%\apj, 433, 583

%\bibitem[Woosley 1987]{W87}
%Woosley, S.E. 1987,
%Nucleosynthesis and Chemical Evolution,
%ed. B. Hauck \etal, Geneva: Geneva Obs., 1

%\bibitem[Woosley 1995]{Woos95}
%Woosley, S.E. 1995,
%private communication

\bibitem[Woosley \& Weaver 1995]{WW95}
Woosley, S.E. \& Weaver, T.A. 1995,
\apjs, 101, 181  (WW95)

%\bibitem[Woosley \etal 1993]{WLW93}
%Woosley, S.E., Langer, N. \& Weaver, T.A. 1993,
%\apj, 411, 823

\bibitem[Worthey 1994]{W94}
Worthey, G. 1994,
\apjs, 95, 107

\bibitem[Worthey 1995]{W95}
Worthey, G. 1995,
private communication

\bibitem[Worthey \etal 1992]{WFG92}
Worthey, G., Faber, S.M. \& Gonz\'alez, J.J. 1992,
\apj, 398, 69

\bibitem[Worthey \etal 1995]{WTF95}
Worthey, G., Trager, S.C. \& Faber, S.M. 1995,
Fresh Views on Elliptical Galaxies, ed. A. Buzzoni \& A.
Renzini, ASP Conf. Series, in press

\bibitem[Worthey \etal 1996]{WDJ96}
Worthey, G., Dorman, B. \& Jones, L.A. 1996,
\aj, 112, 948

%\bibitem[Yoshii \& Arimoto 1987]{YA87}
%Yoshii, Y. \& Arimoto, N. 1987,
%\aap, 188, 13

\bibitem[Zepf \& Silk 1996]{ZS96}
Zepf, S.E. \& Silk, J. 1996,
\apj, 466, 114

\end{thebibliography}
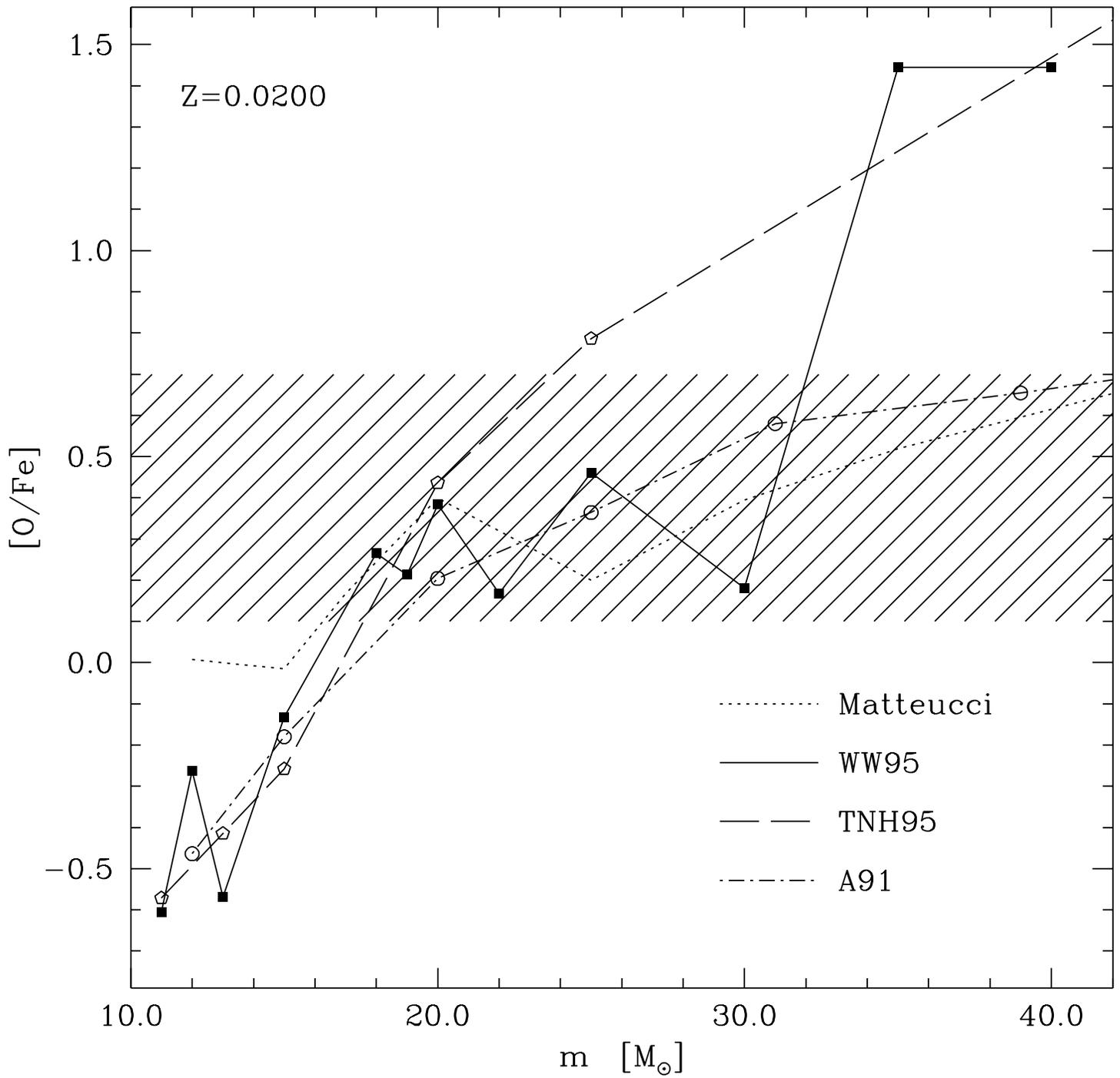
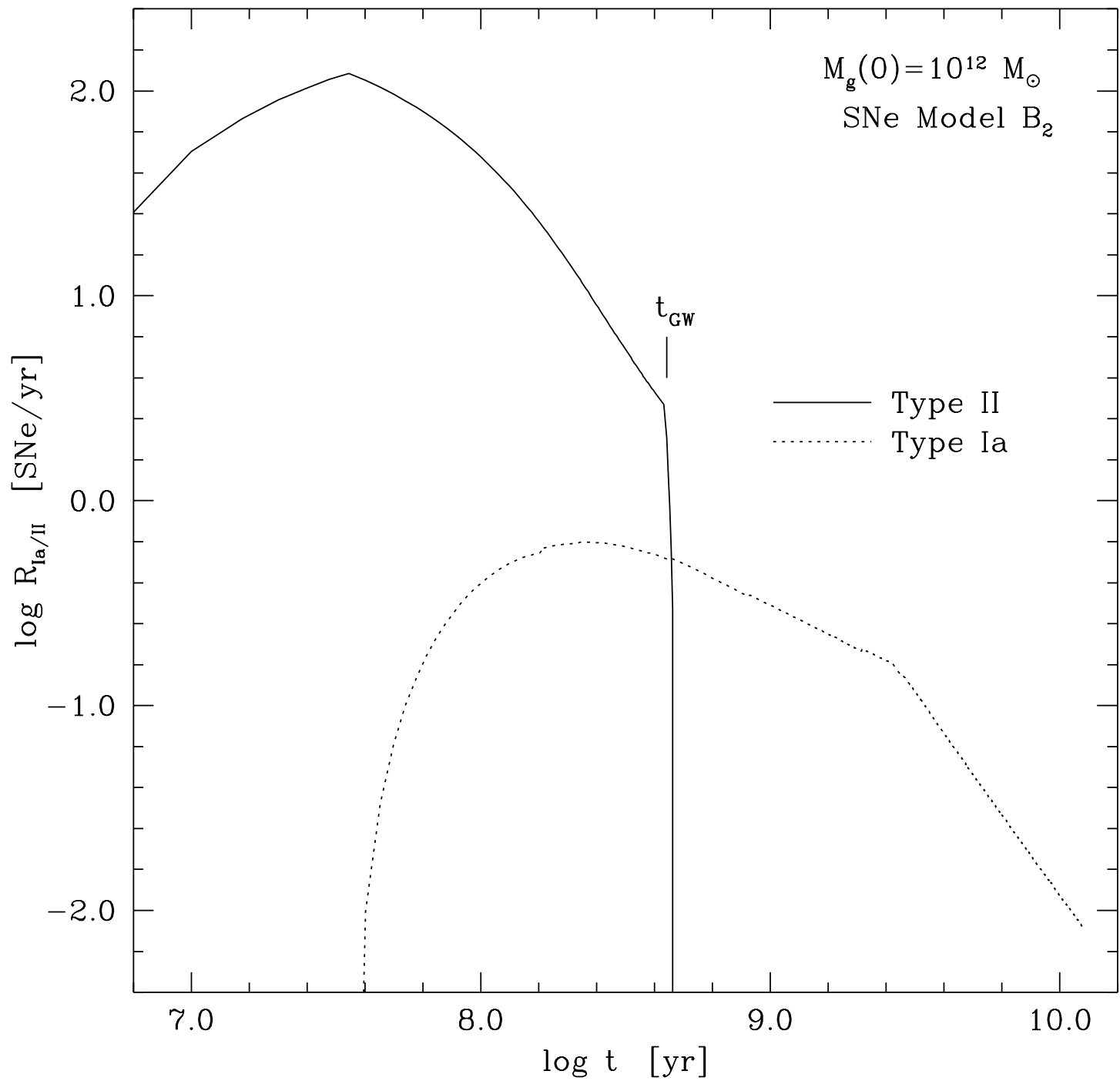
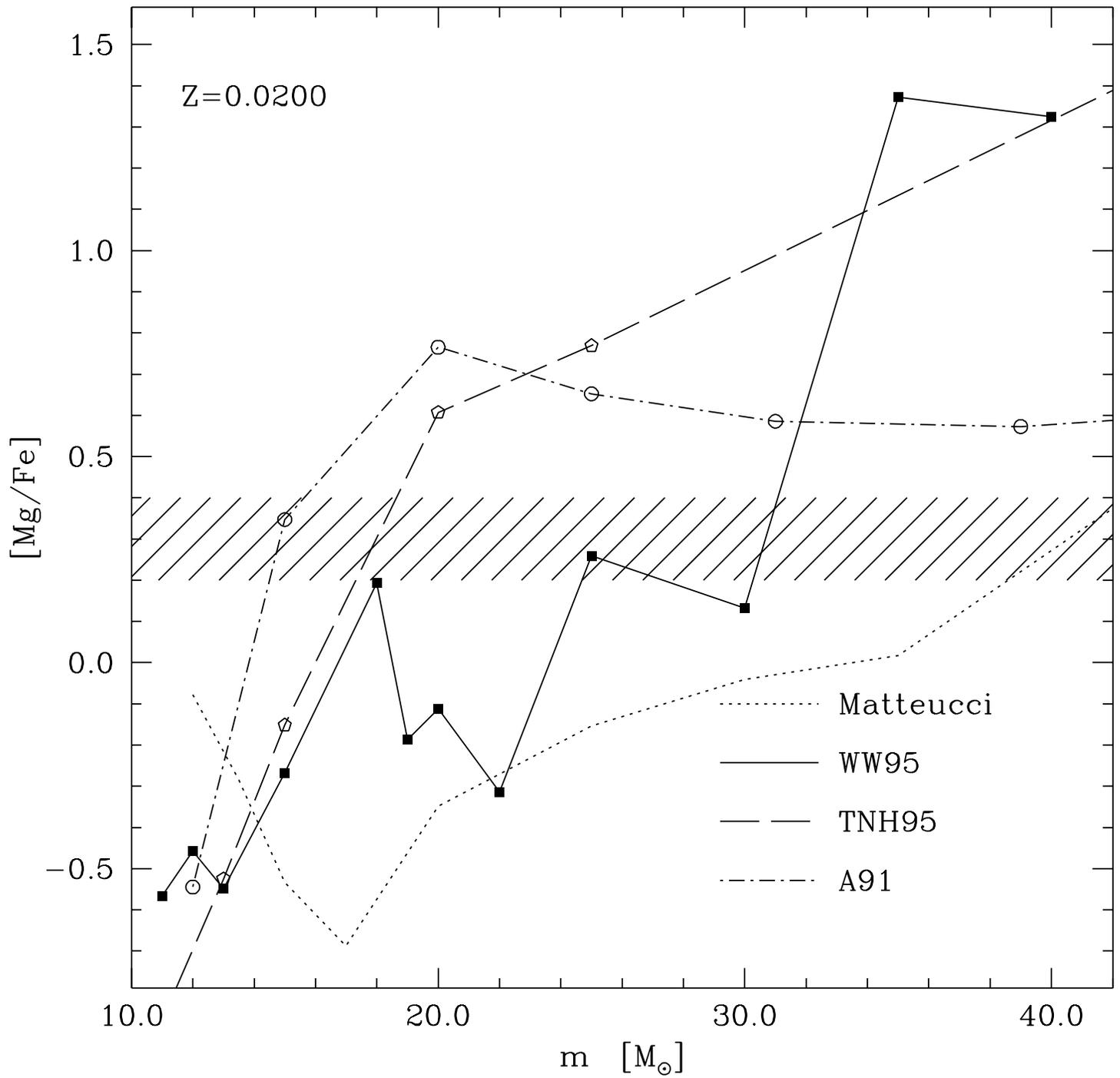
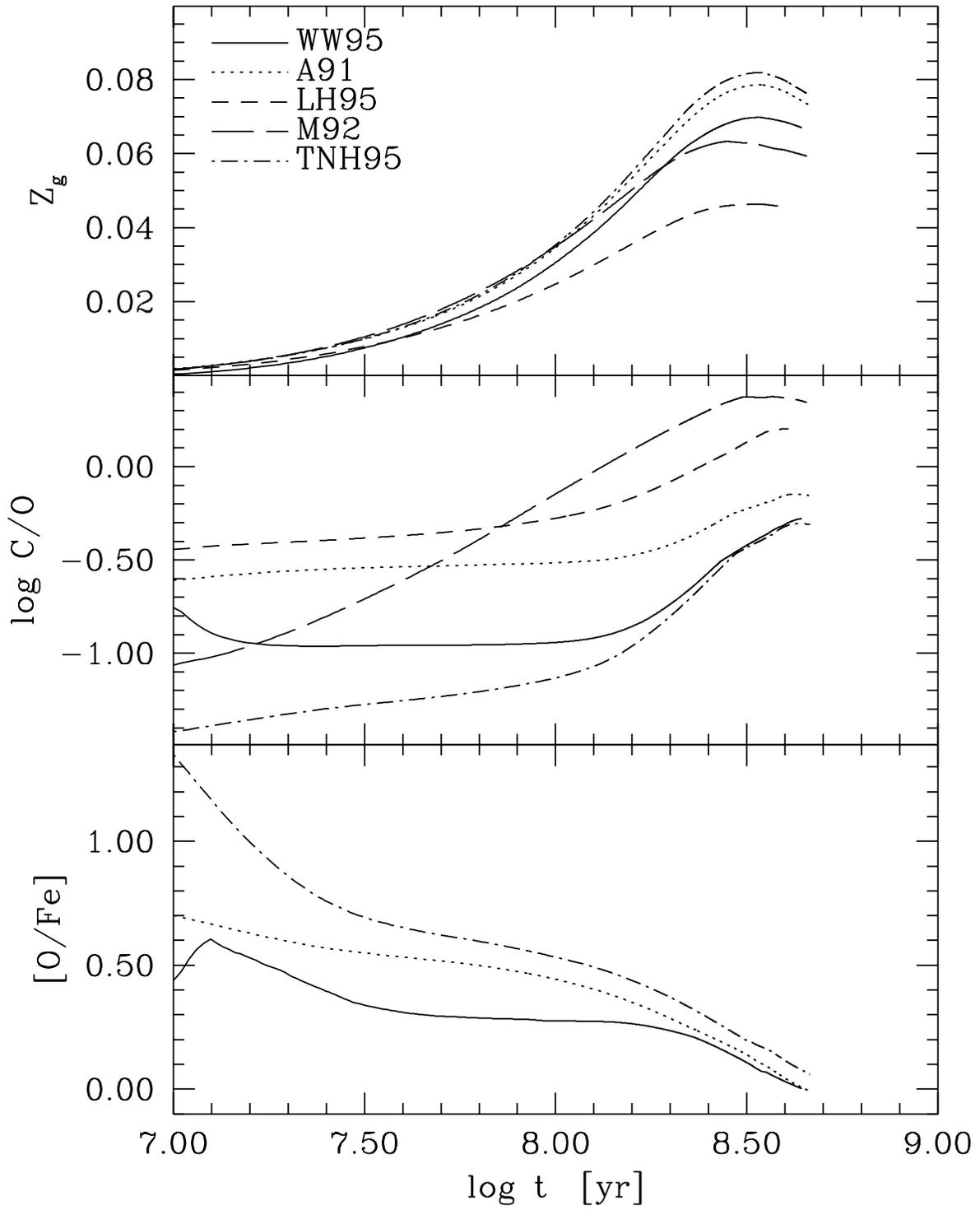
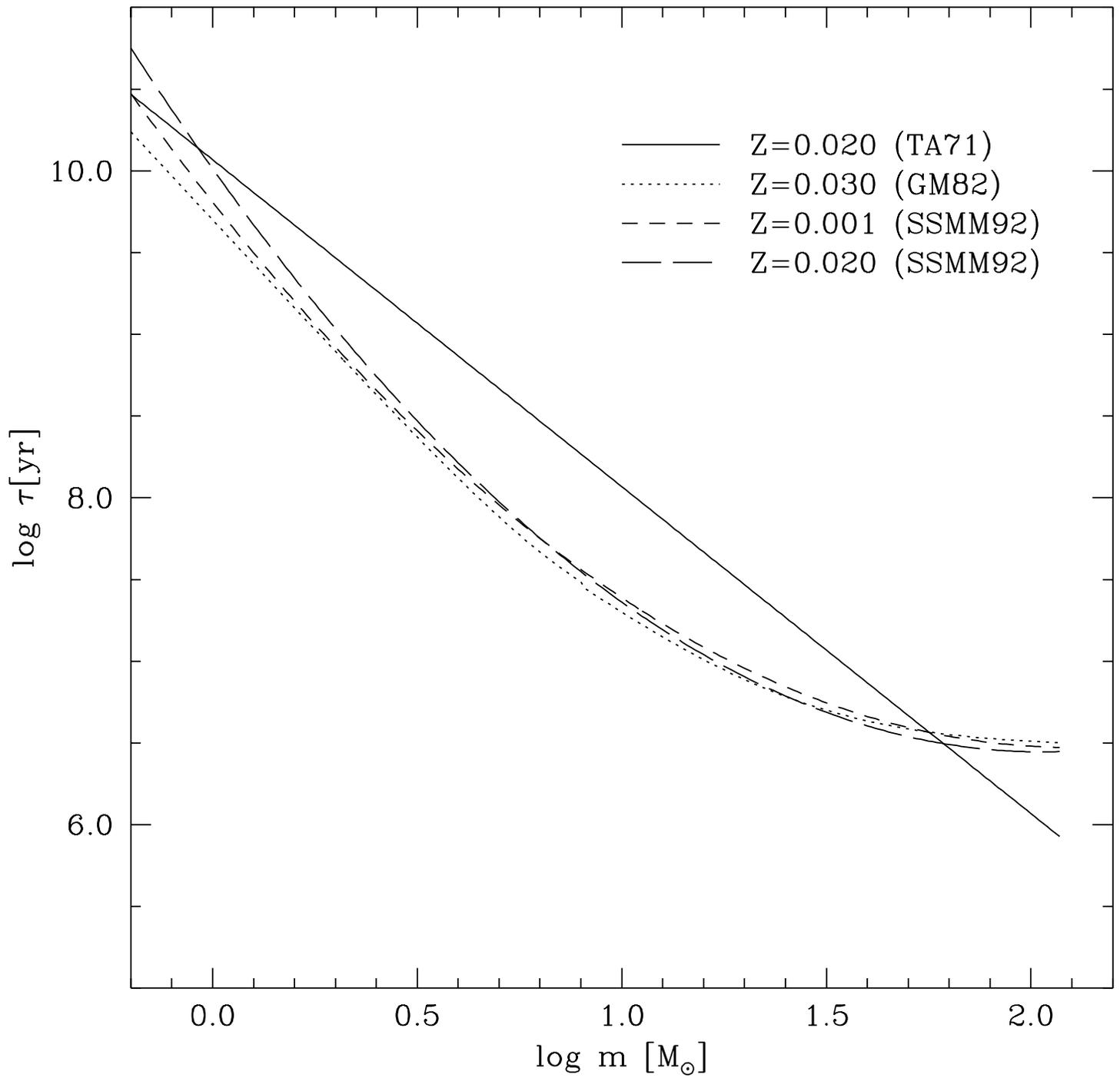
\end{document}